%% file: HIN-24-021_temp.tex
\documentclass[11pt,twoside,a4paper,cmspaper,final,collab]{cms-tdr}
\def\svnVersion{21dd277}\def\svnDate{2026/08/24}\def\cmsCernNoTag{CERN-EP-2026-066}\def\cmsCernDate{\today}\def\cmsMessage{Published in the Journal of High Energy Physics as \href{http://dx.doi.org/10.1007/JHEP08(2026)054}{\doi{10.1007/JHEP08(2026)054}.}}
\begin{document}\cmsNoteHeader{HIN-24-021}

\newlength\cmsTabSkip\setlength{\cmsTabSkip}{1ex}

\newcommand{\Wjets}{\ensuremath{\PW+\text{jets}}\xspace}
\newcommand{\Zjets}{\ensuremath{\PZ+\text{jets}}\xspace}
\newcommand{\Zjet}{\ensuremath{\PZ+\text{jet}}\xspace}
\newcommand{\PPb}{\ensuremath{\mathrm{Pb}}\xspace}
\newcommand{\PbPb}{\ensuremath{\PPb\PPb}\xspace}
\newcommand{\pp}{\ensuremath{\Pp\Pp}\xspace}
\newcommand{\ee}{\ensuremath{\Pep\Pem}\xspace}
\newcommand{\mumu}{\ensuremath{\PGmp\PGmm}\xspace}
\newcommand{\emu}{\ensuremath{\Pemp\PGmpm}\xspace}
\newcommand{\WW}{\ensuremath{\PW\PW}\xspace}
\newcommand{\DY}{\ensuremath{\mathrm{DY}}\xspace}
\newcommand{\RttDY}{\ensuremath{R_{\ttbar/\DY}}\xspace}
\newcommand{\Qlp}{\ensuremath{\Pellp}\xspace}
\newcommand{\Qlm}{\ensuremath{\Pellm}\xspace}
\newcommand{\QAGnl}{\ensuremath{\HepAntiParticle{\PGn}{\Pell}{}}\xspace}
\newcommand{\QGnl}{\ensuremath{\HepParticle{\PGn}{\Pell}{}}\xspace}
\newcommand{\mub}{\unit{\ensuremath{\mu}b}\xspace}
\newcommand{\sigmaTTbarobs}{\ensuremath{3.42 ^{+0.54}_{-0.51} \stat^{+0.50}_{-0.43} \syst}\xspace}
\newcommand{\sigmaDYobs}{\ensuremath{397 \pm 3 \stat^{+27}_{-25} \syst}\xspace}
\newcommand{\RTTbarDYobs}{\ensuremath{0.0086 ^{+0.0014}_{-0.0013} \stat^{+0.0011}_{-0.0010} \syst}\xspace}

\cmsNoteHeader{HIN-24-021}

\title{Measurement of the top quark pair production cross section in \texorpdfstring{\PbPb}{PbPb} collisions at\texorpdfstring{$\sqrtsNN=5.36\TeV$}{sqrt(s_NN) = 5.36 TeV}}

\date{\today}

\abstract{The inclusive cross section for top quark pair (\ttbar) production in lead-lead (\PbPb) collisions is reported for the first time at a center-of-mass energy per nucleon pair of 5.36\TeV. The analysis uses data corresponding to an integrated luminosity of 1.58\nbinv collected by the CMS experiment at the CERN LHC in 2023. The \ttbar production cross section, $\sigma_{\ttbar}=\sigmaTTbarobs\mub$, is measured in dilepton final states using a fit to a multivariate discriminator that combines the decay electron and muon kinematic properties with the multiplicity of bottom quark jets. The result is consistent with perturbative quantum chromodynamics calculations at next-to-next-to-leading order (NNLO) accuracy employing several nuclear parton distribution functions. In addition, the Drell--Yan production cross section ($\sigma_\text{DY}$) for dilepton masses above 10\GeV and the ratio of \ttbar to DY cross sections ($\RttDY$) are found to be compatible with the NNLO predictions. The observables $\sigma_{\ttbar}$, $\sigma_\text{DY}$, and $\RttDY$ are measured separately for central and semicentral \PbPb collisions to investigate for the first time the dependence of top quark production on the collision impact parameter.}

\hypersetup{%
pdfauthor={CMS Collaboration},%
pdftitle={Measurement of the top quark pair production cross section in PbPb collisions at sqrt(sNN) = 5.36 TeV},%
pdfsubject={CMS},%
pdfkeywords={CMS, heavy ions, top quark, Run 3}}

\maketitle

\section{Introduction}

Extensive studies of the cross section and properties of top quark-antiquark pairs (\ttbar) produced in
proton-proton (\pp) collisions at the CERN LHC~\cite{CMS:2024gzs,ATLAS:2024kxj} have improved our knowledge
of key fundamental parameters of the standard model, such as the top quark mass ($m_\PQt$),
the strong coupling constant (\alpS), and the Yukawa coupling of the top quark to the Higgs boson, through comparisons
of the data with high-accuracy perturbative quantum chromodynamics (pQCD) predictions. These measurements
have also contributed to determining the parton distribution functions (PDFs) of the proton with reduced
uncertainties. In heavy ion collisions, the production of top quarks has long been proposed as a powerful
probe of the nuclear gluon density at high virtualities~\cite{dEnterria:2015mgr}, as encoded in nuclear
PDFs (nPDFs). It has also been proposed as a useful means to study final-state effects on top quark decay products as they
traverse the produced quark-gluon plasma (QGP), where they may experience parton energy loss in the medium (``jet quenching'')~\cite{dEnterria:2015mgr,Apolinario:2017sob}.

Because the top quark decays almost exclusively as $\PQt\to\PW\PQb$~\cite{ParticleDataGroup:2024cfk}, a \ttbar event contains two \PQb jets and the decay products of the two \PW bosons. Experimentally, \ttbar events are classified according to the \PW boson decay modes.
About 46\% of \ttbar events are fully hadronic, but these are difficult to isolate from quantum chromodynamics (QCD) multijet backgrounds, which have much larger cross sections. Events in which only one \PW boson decays leptonically account for roughly 45\% of the total; however, their final states contain multiple jets and are subject to systematic uncertainties from jet energy calibration, quark flavor tagging, and combinatorial ambiguity. The dilepton decay mode, $\ttbar \to(\PWp\PQb)(\PWm\PAQb) \to(\Qlp\QGnl\PQb)(\Qlm\QAGnl\PAQb)$, provides the cleanest experimental signature, even though its branching fraction is about 4\% when restricting $\Pell$ to electrons and muons. This channel yields a pure (${>}95\%$) sample with relatively loose lepton selection requirements. In $\pp$ collisions, the single lepton~\cite{CMS:2021vhb} and dilepton~\cite{CMS:2018fks} channels provide the most precise measurements of the \ttbar production cross section. Among all channels, the \emu final state typically provides the best sensitivity. Owing to its higher purity and nearly twice the event yields of the same flavor ($\ee$ or $\mumu$) channels, the \emu final state is used to measure the \ttbar production cross section with lower uncertainties. In the heavy ion environment, the presence of a very large hadronic background makes it more challenging to measure the lepton$\,+\,$jets decay final states. For these reasons, the dilepton final state, including events with tau leptons that decay into an electron or a muon, is commonly adopted to measure the \ttbar production cross sections.

The production of top quark pairs has been measured in proton-lead collisions at a nucleon-nucleon ($\PN\PN$) center-of-mass energy of $\sqrtsNN=8.16\TeV$~\cite{CMS:2017hnw,ATLAS:2024qdu} and in lead-lead (\PbPb) collisions at $\sqrtsNN=5.02\TeV$ \cite{CMS:2020aem,ATLAS:2024jji}. The early \PbPb collision measurements used data corresponding to 1.7--1.9\nbinv of integrated luminosity. They have proved useful to establish the baseline observables and selection criteria in the dilepton channel, confirming the feasibility of top quark measurements in events with very large particle multiplicities. However, the ability of these studies to constrain nPDFs or the properties of QGP is limited by the poor statistical precision of the datasets analyzed.

This paper reports the first measurement of the inclusive production cross section of $\PbPb\to\ttbar$ at $\sqrtsNN=5.36\TeV$, utilizing data corresponding to an integrated luminosity of 1.58\nbinv collected in 2023. The analysis uses advanced algorithms to select leptons and identify jets stemming from the fragmentation and hadronization of \PQb quarks (\PQb jets). These improved techniques combine new variables used to characterize the isolation and prompt production of leptons from \ttbar events, and make use of new discriminants based on machine learning to identify the decay leptons, as well as the \PQb jets. By incorporating these novel algorithms and performing a combined fit to a multivariate \ttbar discriminant across several event categories, the analysis improves upon previous results, reducing the uncertainty in the $\sigma(\PbPb\to\ttbar)$ cross section and enabling the first observation of this process in heavy ion collisions by the CMS experiment.

The paper is structured as follows. The experimental apparatus is presented in Section~\ref{sec:cms}, and the simulation and event selections are described in Sections~\ref{sec:MC} and~\ref{sec:dataandsel}, respectively. The methods used to control the backgrounds and to extract the signal are described in Sections~\ref{sec:backgrounds} and~\ref{sec:statana}, respectively. The results are presented in Section~\ref{sec:results}, and summarized in Section~\ref{sec:conclusions}. Tabulated results of the analysis can be found in HEPData~\cite{hepdata}.

\section{The CMS detector}
\label{sec:cms}

The CMS apparatus~\cite{CMS:2008xjf,CMS:2023gfb} is a multipurpose, nearly hermetic detector, designed to trigger
on~\cite{CMS:2020cmk,CMS:2016ngn,CMS:2024aqx} and identify electrons, muons, photons, and (charged and neutral)
hadrons~\cite{CMS:2020uim,CMS:2018rym,CMS:2014pgm}. Its central feature is a superconducting solenoid of 6\unit{m}
internal diameter, providing a magnetic field of 3.8\unit{T}. Within the solenoid volume are a silicon pixel and strip tracker,
a lead tungstate crystal electromagnetic calorimeter (ECAL), and a brass and scintillator hadron calorimeter (HCAL),
each composed of a barrel and two endcap sections.
Muons are reconstructed using gas-ionization detectors embedded in the steel flux-return yoke outside the solenoid.
Forward calorimeters extend the pseudorapidity ($\eta$) coverage provided by the barrel and endcap detectors.
Two forward hadron (HF) calorimeters use steel as an absorber and quartz fibers as the sensitive material. They are located 11.2\unit{m} from the interaction region, one on each end, and together they provide coverage in the range $3.0 < \abs{\eta} < 5.2$.
They also serve as luminosity monitors. In addition, two zero-degree calorimeters (ZDCs), made of quartz fibers and plates
embedded in tungsten absorbers, are used to detect neutrons (\Pn) from nuclear breakup events in the range $\abs{\eta} > 8.3$~\cite{Grachov:2006ke}.
More detailed descriptions of the CMS detector, together with a definition of the coordinate system used and the
relevant kinematic variables, can be found in Refs.~\cite{CMS:2008xjf,CMS:2023gfb}.

Events of interest are selected using a two-tiered trigger system. The first level, composed of custom hardware processors,
uses information from the calorimeters and muon detectors to select events at a rate of up to 70\unit{kHz} within a
fixed latency of 4\mus~\cite{CMS:2020cmk}. The second level, known as the high-level trigger (HLT), consists of a farm of
processors running a version of the full event reconstruction software optimized for fast processing, and reduces the
event rate down to 50\unit{kHz} before data storage~\cite{CMS:2016ngn,CMS:2024aqx}. The quoted values have been optimized for \PbPb collisions.

A particle-flow (PF) algorithm~\cite{CMS:2017yfk} aims to reconstruct and identify each individual particle in an event,
with an optimized combination of information from the various elements of the CMS detector. The energy of photons is
obtained from the ECAL measurement. The energy of an electron is determined from a combination of the electron momentum
at the primary interaction vertex as determined by the tracker, the energy of the corresponding ECAL cluster, and the
energy sum of all bremsstrahlung photons spatially compatible with originating from the electron track.
The energy of a muon is obtained from the curvature of the corresponding track. The energy of a charged hadron is
determined from a combination of the momentum measured in the tracker and the matching ECAL and HCAL energy deposits,
corrected for the response function of the calorimeters to hadronic showers. Finally, the energy of neutral hadrons is
obtained from the corresponding corrected ECAL and HCAL energies.

\section{Monte Carlo simulations}
\label{sec:MC}

Monte Carlo (MC) simulations are used to estimate the signal acceptance and efficiency, as well as to validate the background estimation
methods. The simulations are performed using matrix element (ME) generators based on pQCD calculations at next-to-leading order (NLO) accuracy, emulating a mixture of \pp, \Pp{}\Pn, and \Pn{}\Pn\ interactions in proportions corresponding to their relative abundances in
lead nuclei (mass number $A=208$). The ME generators employ the EPPS21\_nlo nPDFs~\cite{Eskola:2021nhw}, which use the CT18ANLO PDF~\cite{Hou:2019efy} as the proton parton density baseline. The generated events are then interfaced with the \PYTHIA~8.306~\cite{Sjostrand:2014zea} parton shower (with the CP5 tune~\cite{CMS:2019csb}) and are embedded in inelastic \PbPb events simulated with the \HYDJET~\cite{Lokhtin:2008xi} MC generator. The \HYDJET model reproduces the effects of the underlying event (UE) measured in \PbPb\ data.
In doing so it does not alter the hard-scattering process simulated at the ME level, nor the decay products of the generated particles. To account for possible medium effects affecting the simulated signal and background final states, methods based on data are employed and described in more detail at the end of Section~\ref{sec:dataandsel}. The \PbPb collision ``centrality'' (\ie, the degree of overlap of the two \PPb ions) is determined from the sum of the transverse energy ($\sum\ET$, where $\ET=E\cosh\eta$), deposited in both HF calorimeters and corrected with an event-by-event weighting procedure based on the average number of $\PN\PN$ collisions in each centrality interval as obtained with a Glauber model~\cite{dEnterria:2020dwq}. After this weighting, residual differences between the centrality distribution measured in data and predicted in simulation are corrected for based on a control sample of $\PZ\to\mumu$ events. All the MC samples include a simulation of the full detector response implemented in \GEANTfour~\cite{Agostinelli:2002hh} with realistic alignment and calibration, tuned to data.

{\tolerance=800 The MC \ttbar signal samples are generated at NLO accuracy with \MGvATNLO (v2.9.18)~\cite{Alwall:2014hca} and \POWHEG v2~\cite{Frixione:2007nw}. In the former case, the event generation includes up to two extra partons at the ME level, and the merging with the parton shower is performed via the FxFx procedure~\cite{Frederix:2012ps}. In the \POWHEG case, only up to one real parton emission is generated, with the remainder produced by the parton shower. The cross sections of both simulated \ttbar samples are scaled up to match the next-to-next-to-leading order (NNLO) pQCD predictions, including soft-gluon resummation at next-to-next-to-leading logarithmic (NNLL) accuracy, obtained with the \TOPpp program~\cite{Czakon:2013goa}. In the calculation a top quark mass of $m_\PQt = 172.5\GeV$ is used, the nominal renormalization and factorization scales are set to $\mu_\text{R} = \mu_\text{F} = m_\PQt$, and the QCD strong coupling is set to $\alpS(m_\PZ)=0.118$. Uncertainties of ${\pm}1.0\GeV$ and ${\pm}0.001$ are assumed for the $m_\PQt$ and $\alpS(m_\PZ)$ values~\cite{ParticleDataGroup:2024cfk}, respectively. The predicted NNLO$\,+\,$NNLL inclusive cross section is $\sigma_\text{th}(\PbPb\to\ttbar) = 3.69^{+0.71}_{-0.48}\mub$, where the uncertainty is the quadratic sum of theoretical scale, parametric (from $m_\PQt$ and $\alpS$), and EPPS21\_nlo nPDF uncertainties.\par}

The characteristic signature of the \ttbar events selected in this analysis is the presence of two leptons with large transverse momentum ($\pt$) originating from the $\PWp\PWm$ decays. Consequently, any process that produces a similar dilepton final state can mimic the signal and constitutes a background to the measurement. The simulated background samples include Drell--Yan (\DY, $\PZ/\Pgg^*\to\ee/\mumu/\PGtp\PGtm$), \Wjets, single top quark production, diboson processes, and multijet events containing nonprompt leptons, mostly from heavy-quark hadron decays. The inclusive DY process $\pp\to\PZ+\PX$, including up to two recoiling jets via the FxFx merging procedure, is simulated with \MGvATNLO separately in the low-mass (10--50\GeV) and high-mass (${>}50\GeV$) regions, with predicted cross sections of 287 and $83\mub$, respectively. The same generator is used for the \Wjets sample, with a cross section of 899\mub. Single top quark production in association with a \PW boson or via the $t$ channel is simulated with \POWHEG~\cite{Re:2010bp}, with corresponding cross sections of 0.272 and $1.27\mub$. Diboson events, in particular \WW, are generated with \POWHEG including all final states with at least one lepton. The \WW contribution is normalized to a cross section of $0.613\mub$, as predicted by \POWHEG. Finally, multijet processes with at least one muon or electron are simulated with \PYTHIA. The multijet samples are included in the training of the lepton isolation discriminators, but are not used to estimate any of the backgrounds.

\section{Event selection}
\label{sec:dataandsel}

Candidate events are first selected based on whether they pass the minimum-bias or the single-muon HLT triggers. The minimum-bias trigger requires a coincident signal in both HF calorimeters and the presence of at least one neutron in one of the ZDCs, whereas the single-muon trigger requires a muon reconstructed at the HLT with $\pt>7\GeV$. The stability of these triggers is verified throughout the data-taking period, and their efficiency is extracted directly from the data. Offline, collision events are further required to have at least one interaction vertex reconstructed with two or more tracks within a distance of 25\cm from the nominal collision point along the beam axis and 2\cm in the transverse plane. In addition, at least two towers (arrays of channels) on each side of the HF calorimeter must have a total reconstructed energy larger than 4\GeV.

Events containing at least two charged leptons---where ``lepton'' (denoted by $\Pell$) refers exclusively to electrons and muons hereafter---are selected, with additional criteria applied to target dilepton decays of the two \PW\ bosons from \ttbar\ production. Lepton candidates must have $\pt>20\GeV$, originate promptly from the primary collision vertex, and be isolated from other particles in the event. Electrons are required to lie within the ECAL barrel or endcap fiducial regions ($\abs{\eta}\in[0,1.444]\cup [1.566,2.5]$), and muons are selected within $\abs{\eta}<2.4$. Candidates passing these kinematic and fiducial selections are further considered in the analysis if they satisfy two distinct sets of identification and isolation criteria. Decoupling these two requirements enables the isolation variable to define control regions (CRs), which are used to estimate the nonprompt-lepton background and extrapolate it to the signal region.

Electron identification relies on a dedicated boosted decision tree (BDT) discriminant that combines several observables, including the shower shape and position of the ECAL cluster, its matching to the associated reconstructed track, and the energy deposits in the HCAL towers subtended by the ECAL cluster. The BDT is trained on QCD multijet and \ttbar simulations using \textsc{XGBOOST}~\cite{DBLP:journals/corr/ChenG16} and optimized with a grid-search cross-validation procedure~\cite{DBLP:journals/corr/abs-1201-0490}. An electron candidate is considered to pass the identification requirement if the BDT discriminant exceeds a centrality-dependent threshold corresponding to a 95\% efficiency and an expected misidentification rate below 4\%. This procedure ensures a selection efficiency independent of the collision centrality. The contamination from misidentified electron candidates is significantly reduced with respect to the rectangular-based identification criteria used in our previous study~\cite{CMS:2020aem}. Muon identification is based on a set of quality requirements applied to the global track fit, which combines information from the inner tracker and the muon system. The selection follows the standard ``tight'' criterion described in Ref.~\cite{CMS:2018rym}.

For the isolation criteria, a common procedure applicable to both electrons and muons has been further developed, making use of dedicated BDTs trained separately for each lepton type. These BDTs combine different isolation sums of PF candidate energies within cones of radius $R=\sqrt{\smash[b]{\left(\Delta\eta\right)^2+\left(\Delta\phi\right)^2}}=0.3$ around the lepton direction. These isolation sums are computed using the ``soft killer'' method~\cite{Cacciari:2014gra} with thresholds estimated in binned ranges of detector pseudorapidity, namely $[-2.5,\,-2,\,-0.8,\,0.8,\,2,\,2.5]$. This procedure effectively mitigates the dependence of the isolation sums on the \PbPb event activity and enables a robust multivariate discrimination, less reliant on the simulation model used. In addition, the isolation BDT makes use of the three-dimensional impact parameter significance of each lepton. This variable is largely uncorrelated with the isolation sums and provides good discrimination against leptons from heavy-quark hadron decays. As for the electron identification BDT case, the independence of the working point on centrality ensures that the isolation requirement has a constant efficiency of 90\%.

In case more than two leptons pass the selections above, priority is given to muons over electrons. The efficiencies associated with the trigger, identification, and isolation selections are determined using the tag-and-probe method based on $\PZ\to\Qlp\Qlm$ events in data~\cite{CMS:2020uim,CMS:2018rym}. To improve the identification of the signal, the selected events are further categorized as follows:

\begin{itemize}
\item ``SF/DF'' labels indicate whether the leptons have the same ($\ee$ or $\mumu$) or different (\emu) flavors. For SF final states, it is required that the acoplanarity $A_\phi=1-\abs{\Delta\phi(\Pellp\Pellm)}/\pi$ is larger than 0.01 to remove contributions from elastic two-photon production ($\PGg\PGg\to\Pellp\Pellm$)~\cite{CMS:2024bnt}.
\item ``SS/OS'' labels indicate whether the leptons have the same or opposite electric-charge signs.
\item ``on-Z/off-Z'' labels indicate whether the invariant mass of the SF OS pair is $\abs{m(\Pell\Pell)-91\GeV}<15\GeV$ or $\abs{m(\Pell\Pell)-91\GeV}>15\GeV$. Events with SF OS dilepton invariant mass $m(\Pell\Pell)<20\GeV$ are excluded to remove background contributions from quarkonia decays and the low-mass DY process.
\end{itemize}

To improve the discrimination from the main backgrounds, three additional BDT classifiers are trained for the \ee, \emu, and \mumu final states. The BDTs make use of the following lepton-based variables:
\begin{enumerate}
   \item the \pt of the leading (\ie, highest $\pt$) and subleading leptons: $\pt(\Pell_1)$, $\pt(\Pell_2)$;
   \item the asymmetry of the leptons' \pt: $[\pt(\Pell_1)-\pt(\Pell_2)]/[\pt(\Pell_1)+\pt(\Pell_2)]$;
   \item the \pt and mass of the dilepton system: $\pt(\Pell\Pell)$, $m(\Pell\Pell)$;
   \item the sphericity of the dilepton system: $\pt(\Pell\Pell)/[\pt(\Pell_1)+\pt(\Pell_2)]$;
   \item the absolute pseudorapidity of the dilepton system: $\abs{\eta(\Pell\Pell)}$;
   \item the sum of the lepton pseudorapidities: $\sum\abs{\eta_\Pell}$; and
   \item the acoplanarity $A_\phi$ of the dilepton system.
\end{enumerate}

\begin{figure}[htbp]
\centering
\includegraphics[width=0.49\textwidth]{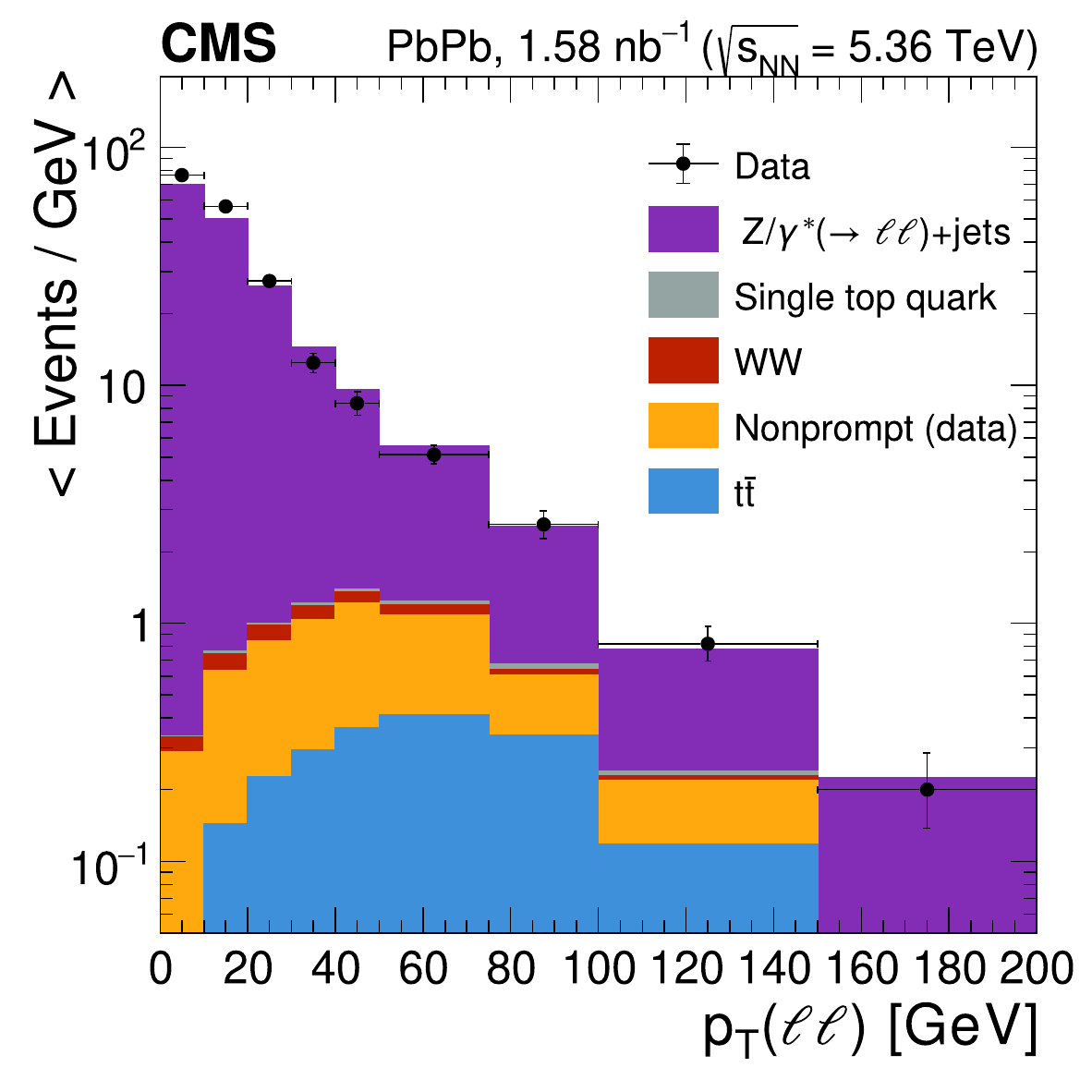}
\includegraphics[width=0.49\textwidth]{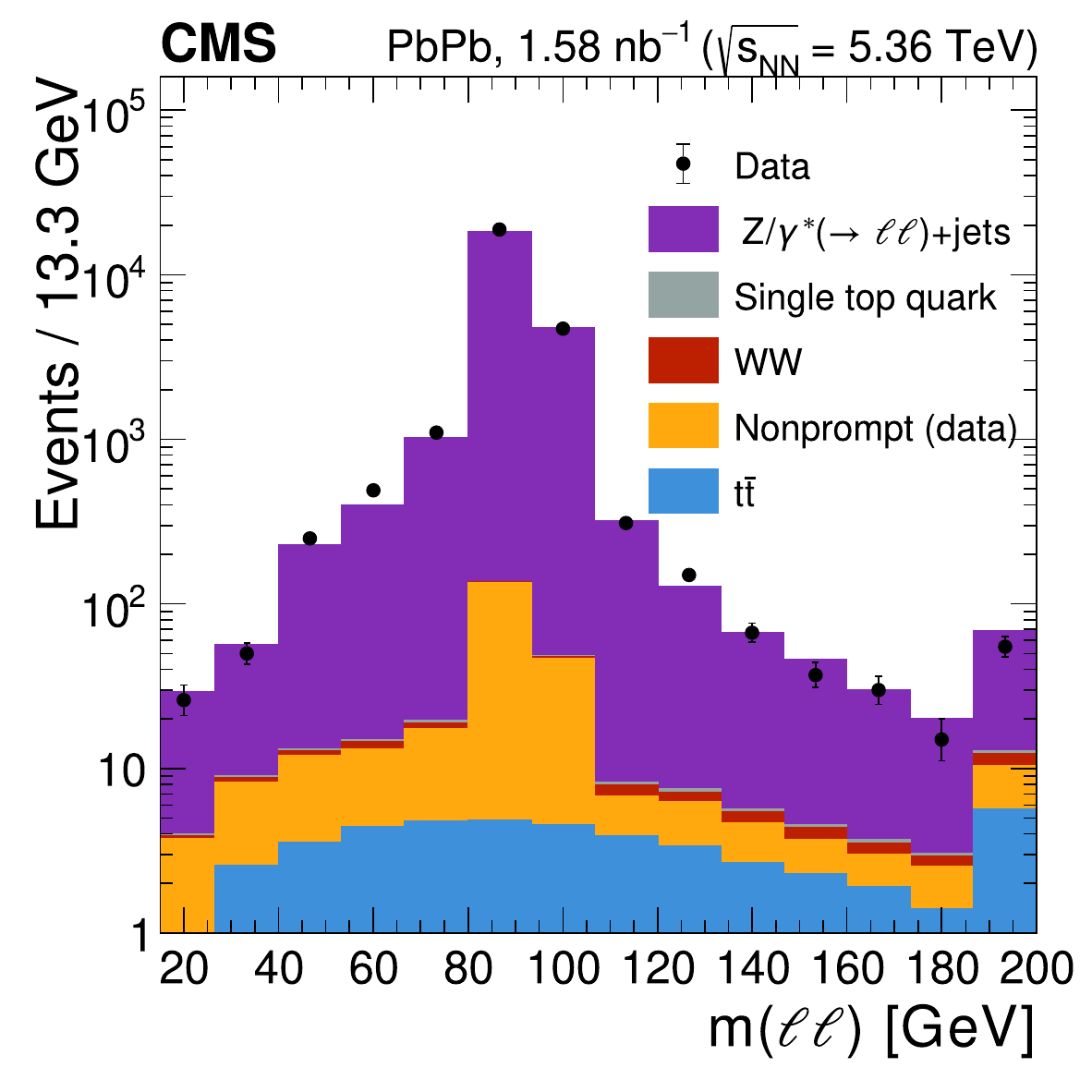}
\caption{Distributions of dilepton $\pt(\Pell\Pell)$ (left) and invariant mass $m(\Pell\Pell)$ (right) variables in the same-flavor ($\Pell\Pell$) channels. The $m(\Pell\Pell)$ distribution is shown before applying the veto of the \PZ boson resonant region. The data (black markers with error bars representing statistical uncertainties) are overlaid on top of the stacked contributions from the expected \ttbar (blue), nonprompt (light orange), \WW (red), single top quark (gray), and \DY (violet) processes. The last bin includes the overflow entries. In the left panel, the event counts are divided by the variable bin width.}
\label{fig:features_sf}
\end{figure}

\begin{figure}[htbp]
\centering
\includegraphics[width=0.49\textwidth]{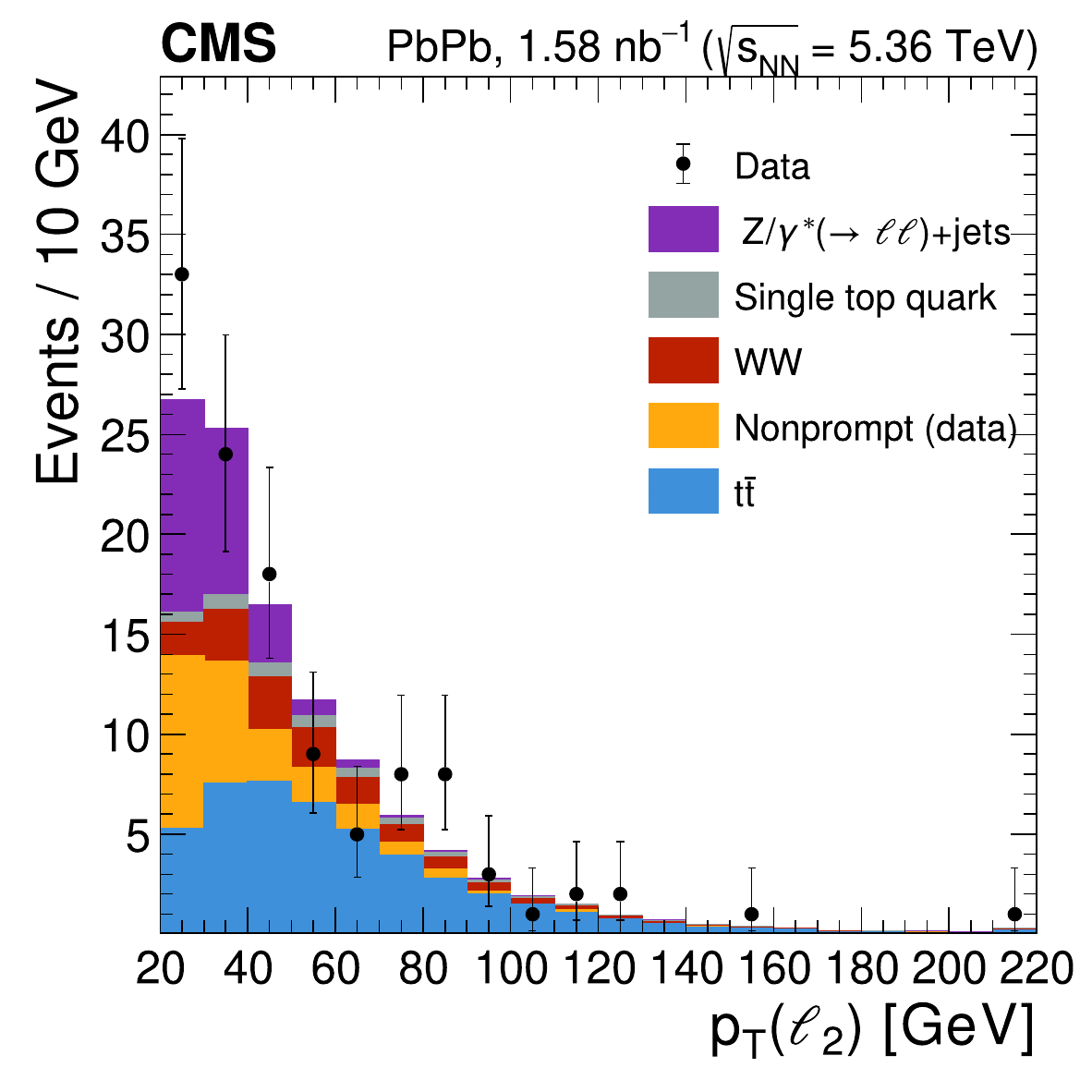}
\includegraphics[width=0.49\textwidth]{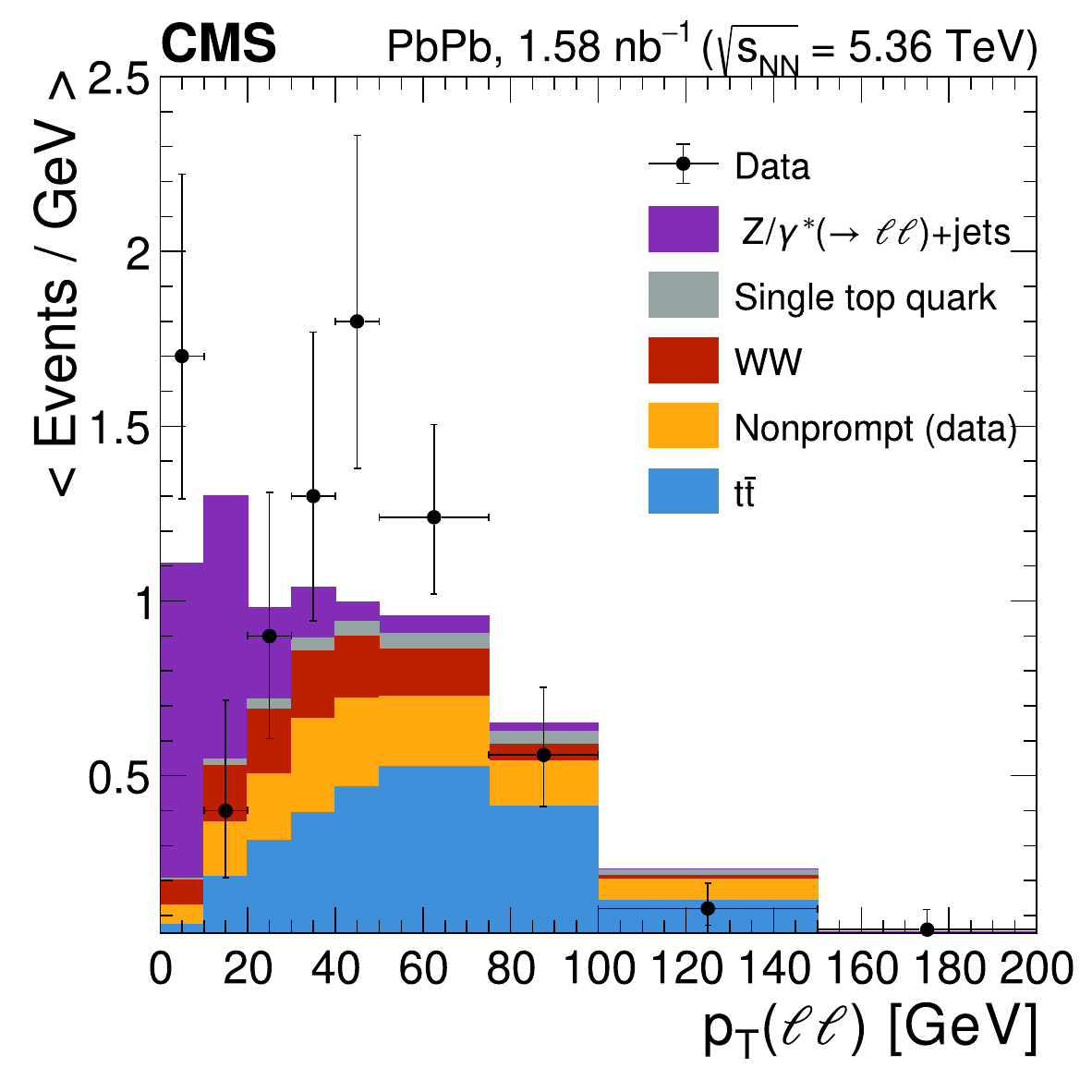}
\includegraphics[width=0.49\textwidth]{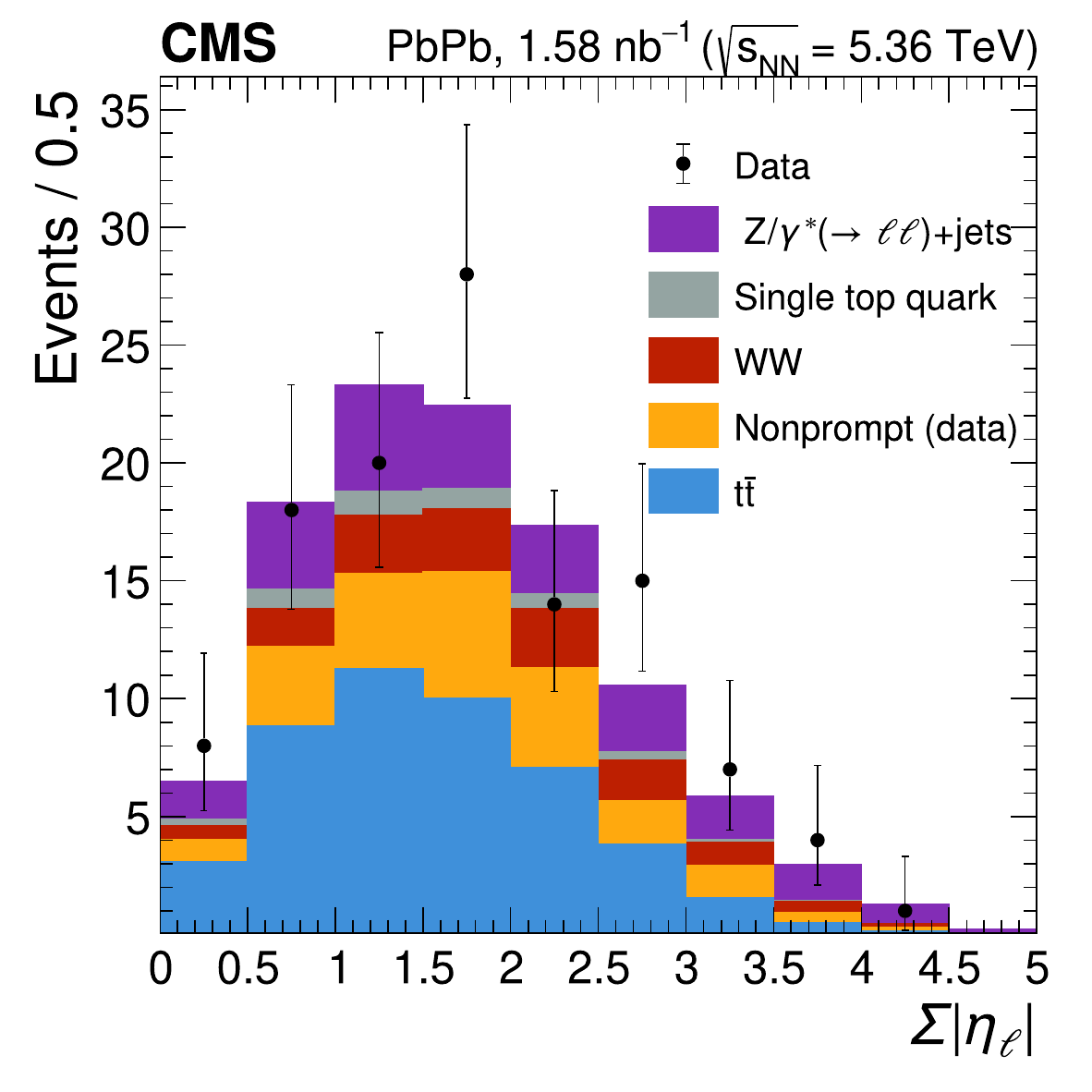}
\caption{
Distributions of $\pt(\Pell_2)$ (upper left), $\pt(\Pell\Pell)$ (upper right), and $\sum\abs{\eta_\Pell}$ (lower) variables in the different-flavor (\emu) channel. The data (black markers with error bars representing statistical uncertainties) are overlaid on top of the stacked contributions from the expected \ttbar (blue), nonprompt (light orange), \WW (red), single top quark (gray), and \DY (violet) processes. The last bin includes the overflow entries.}
\label{fig:features_of}
\end{figure}

Among these, $\pt(\Pell\Pell)$ and $m(\Pell\Pell)$ are found to be the most discriminating variables for the SF channels, whereas $\pt(\Pell_2)$, $\sum\abs{\eta_\Pell}$, and $\pt(\Pell\Pell)$ are the most relevant observables in the DF final state. The relative importance of all input features has been quantified using a \textsc{SHAP} ranking~\cite{Pezoa:2023ulv}. The corresponding distributions are shown in Figs.~\ref{fig:features_sf} and~\ref{fig:features_of} for the same- and different-flavor channels, respectively.
Overall, a good agreement between data and MC expectations is found within the statistical uncertainties, prior to the fit to data. The training of the classifier is performed using the \ttbar (signal) and DY (background) simulations and the \textsc{XGBOOST} package. In the training we have employed a cross-validation procedure and the \textsc{Optuna} algorithm~\cite{optuna_2019} for the optimization of the hyperparameters. In order to improve the sensitivity of the analysis to the \ttbar signal, the BDT discriminator has been evaluated separately for events with zero, one,
or at least two \PQb jets.

For each event, hadronic jets are clustered from the PF candidates using the anti-\kt algorithm~\cite{Cacciari:2008gp, Cacciari:2011ma} with a distance parameter of 0.3. Jet energy corrections and \PQb quark identification are performed with a dedicated multivariate algorithm called ``unified particle transformer'' (\textsc{UParT})~\cite{CMS-DP-2024-066}, explicitly trained for heavy ion collisions.
A detailed description of this algorithm for heavy-flavor quark identification can be found in Refs.~\cite{CMS-DP-2024-088,CMS:2025kje}. Jets with $\pt>30\GeV$, $\abs{\eta}<2.1$, and separated from the lepton by a distance $\Delta R(\Pell,j)>0.4$ are accepted in the analysis. Three working points are defined for the identification of \PQb jets: loose, intermediate, and tight; corresponding to \PQb quark tagging efficiencies of 90, 80, and 70\%, respectively. The intermediate working point is chosen as the default one in the analysis. These working points are parameterized such that the efficiency in simulation is constant as a function of collision centrality. The use of \textsc{UParT} reduces the misidentification rate by more than an order of magnitude relative to the \PQb jet identification algorithm employed during Run~2~\cite{CMS:2020aem}. This improvement stems from the explicit exploitation of the jet PF constituent information within a transformer architecture, in contrast to the Run-2 approach that relied primarily on higher-level variables. In addition, the resulting jet energy scale correction becomes largely independent of the collision centrality (ignoring potential ``jet quenching'' effects) and remains robust with respect to the hadronization model, while retaining a residual dependence on the jet flavor.

\begin{figure}[htbp]
\centering
\includegraphics[width=0.49\textwidth]{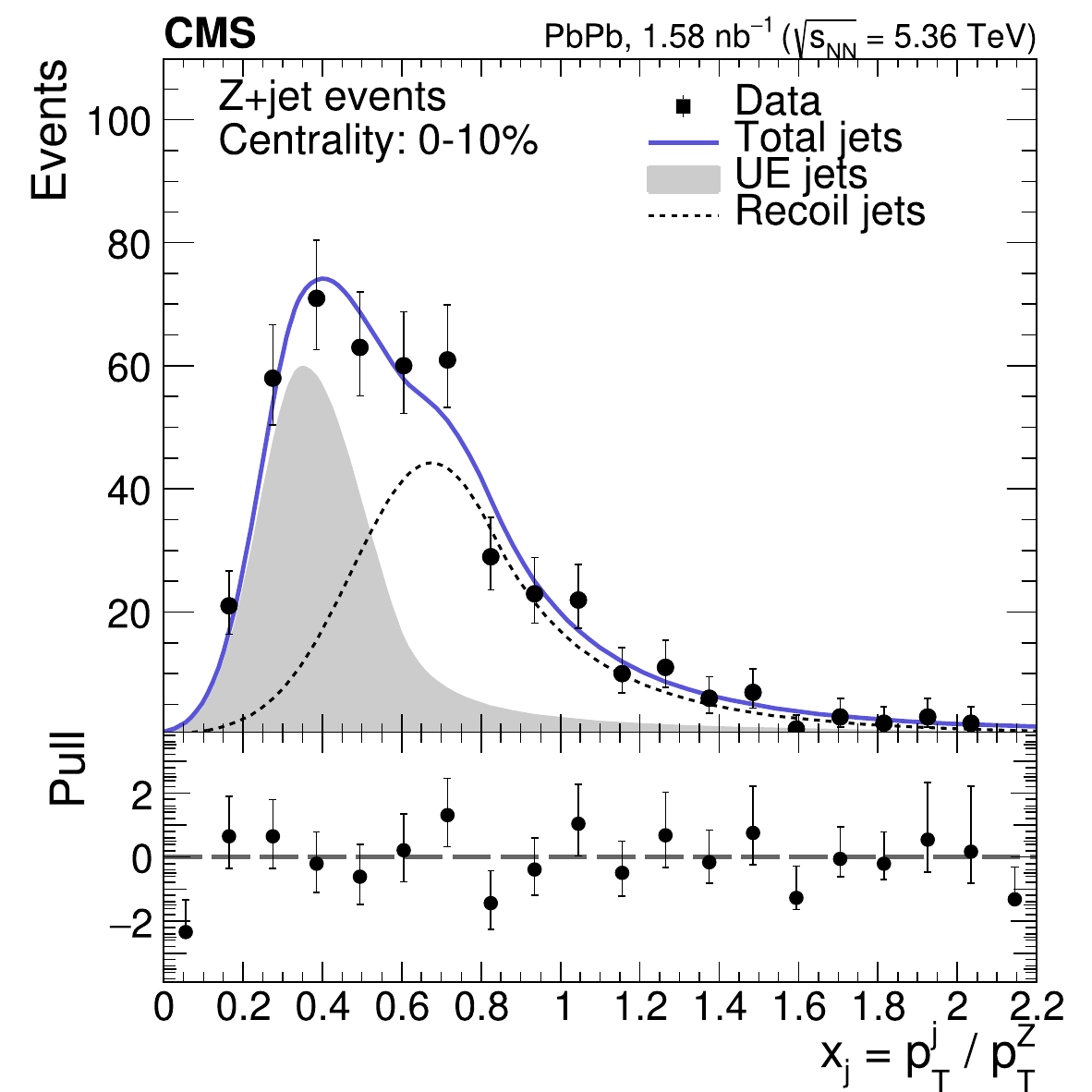}
\includegraphics[width=0.49\textwidth]{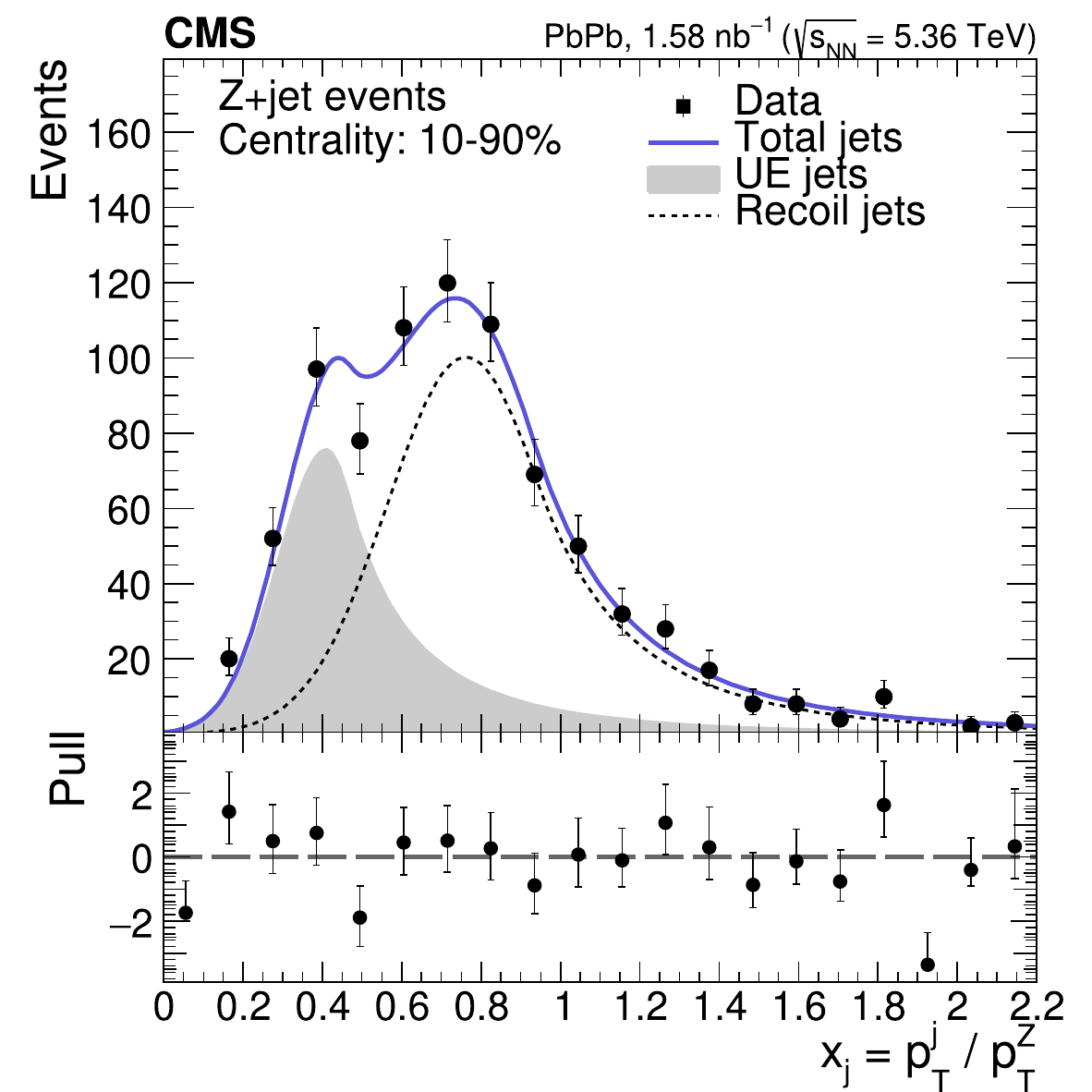}
\caption{Distribution of the $x_\mathrm{j}=\pt(\text{jet})/\pt(\PZ)$ ratio in \Zjet events in the 0--10\% (left) and 10--90\% (right) \PbPb\ centralities. The data (solid dots with error bars) are compared with the contributions from UE jets (gray histograms) and true recoil jets (black dashed curves) stacked. The blue curves show the sum of UE and recoil jet contributions.}
\label{fig:rjetfits_1}
\end{figure}

The residual jet energy scale factor has been studied using \PZ boson events with $\pt(\PZ)>30\GeV$ and a jet recoiling in the transverse plane, $\abs{\Delta\phi(\PZ,\text{jet})}>2\pi/3$. In this control region, the simulation indicates that a significant fraction of the jets originates from genuine recoil contributions, with a smaller component arising from UE jets. Figure~\ref{fig:rjetfits_1} shows the results of fits to the $x_\mathrm{j}=\pt(\text{jet})/\pt(\PZ)$ variable for such $\Zjet$ events in the 0--10\% (left) and 10--90\% (right) \PbPb\ centrality bins. The distribution of recoil jets is modeled using a Crystal Ball function~\cite{Oreglia:1980cs} (black dashed curves), with parameters determined directly from the fit. The UE contribution (gray histograms) dominates in events in which the jet is found in the near-side, or ``toward,'' region relative to the \PZ boson direction, $\abs{\Delta\phi(\PZ,\text{jet})}<\pi/3$. The corresponding $x_\mathrm{j}=\pt(\text{jet})/\pt(\PZ)$ distribution is modeled with the sum of a Crystal Ball and an exponential function. In the fit of $x_\mathrm{j}=\pt(\text{jet})/\pt(\PZ)$, the UE-jets probability density function is fixed and its normalization floats with a Gaussian constraint. The combined recoil$\,+\,$UE-jets model (blue solid curves) reproduces well the data.

\begin{figure}[htbp]
\centering
\includegraphics[width=0.6\textwidth]{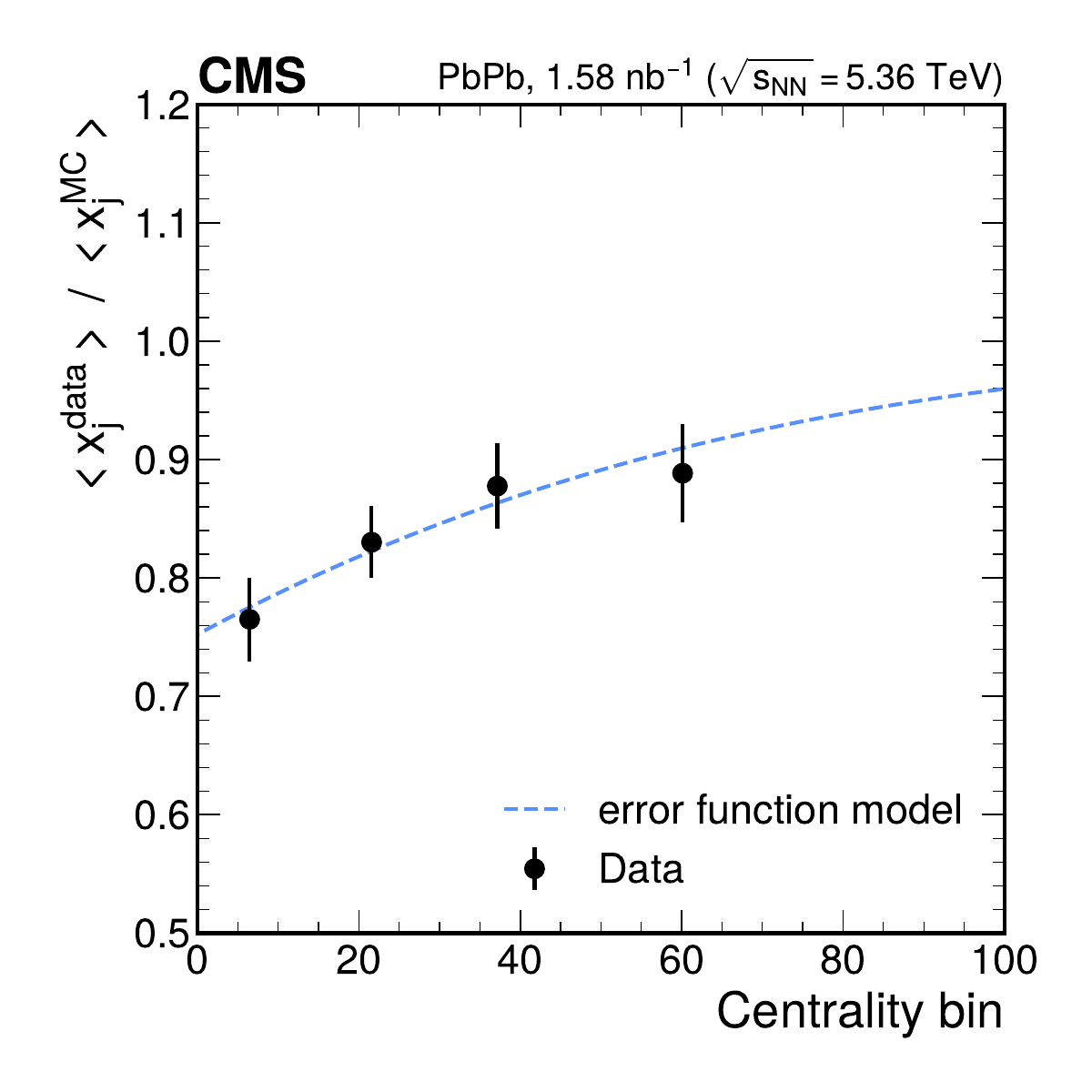}
\caption{Ratio of the average $\langle{x_\mathrm{j}}\rangle$ detector-level distributions measured in data over simulation as a function of the collision centrality with a fit to an error function (blue dashed curve) superimposed.}
\label{fig:rjetfits_2}
\end{figure}

Among all the parameters of the fit, the mean of the Crystal-Ball function is used as a measurement of the average jet energy response, $\langle x_\mathrm{j}\rangle$. The $\langle x_\mathrm{j}\rangle$ profile provides a measure of the residual jet energy response dependence on the collision centrality (Fig.~\ref{fig:rjetfits_2}), and a direct estimate from data of the average jet-quenching effect at detector level. The extrapolation to peripheral collisions ($\text{centrality}>60\%$) shows that, in the absence of final-state medium effects, the data-to-simulation ratio is compatible with unity, indicating good agreement of the simulation with data. For central collisions ($\text{centrality}<10\%$), this ratio is reduced by ${>}20\%$, given that medium effects are absent in the simulation.

The training of \textsc{UParT} has been performed considering separate classes for UE jets and jets from the hard process. This results in an inefficiency of the algorithm in selecting UE jets. Using \Zjet events, the \textsc{UParT} efficiencies have been measured separately for UE and jets recoiling against the \PZ boson. In the regime considered in the analysis, \textsc{UParT} is fully efficient for recoil jets, while being inefficient for UE jets with $\pt<70\GeV$. Data-to-MC scale factors for these efficiencies are parameterized as functions of the collision centrality and the uncorrected jet \pt, and are applied to the baseline simulation as event weights to more accurately reproduce the \textsc{UParT} performance observed in data. The scale factors account for up to 20\% differences between data and MC simulations, being larger for UE jets and in the 0--10\% centrality bin. In the analysis, the simulated jet \pt is additionally corrected using the measured $\langle x_\mathrm{j}\rangle$ profile. The uncertainties associated with this procedure range from 2 to 20\%, depending on the centrality and \pt bins and jet origin.

\begin{figure}[htbp]
\centering
\includegraphics[width=0.49\textwidth]{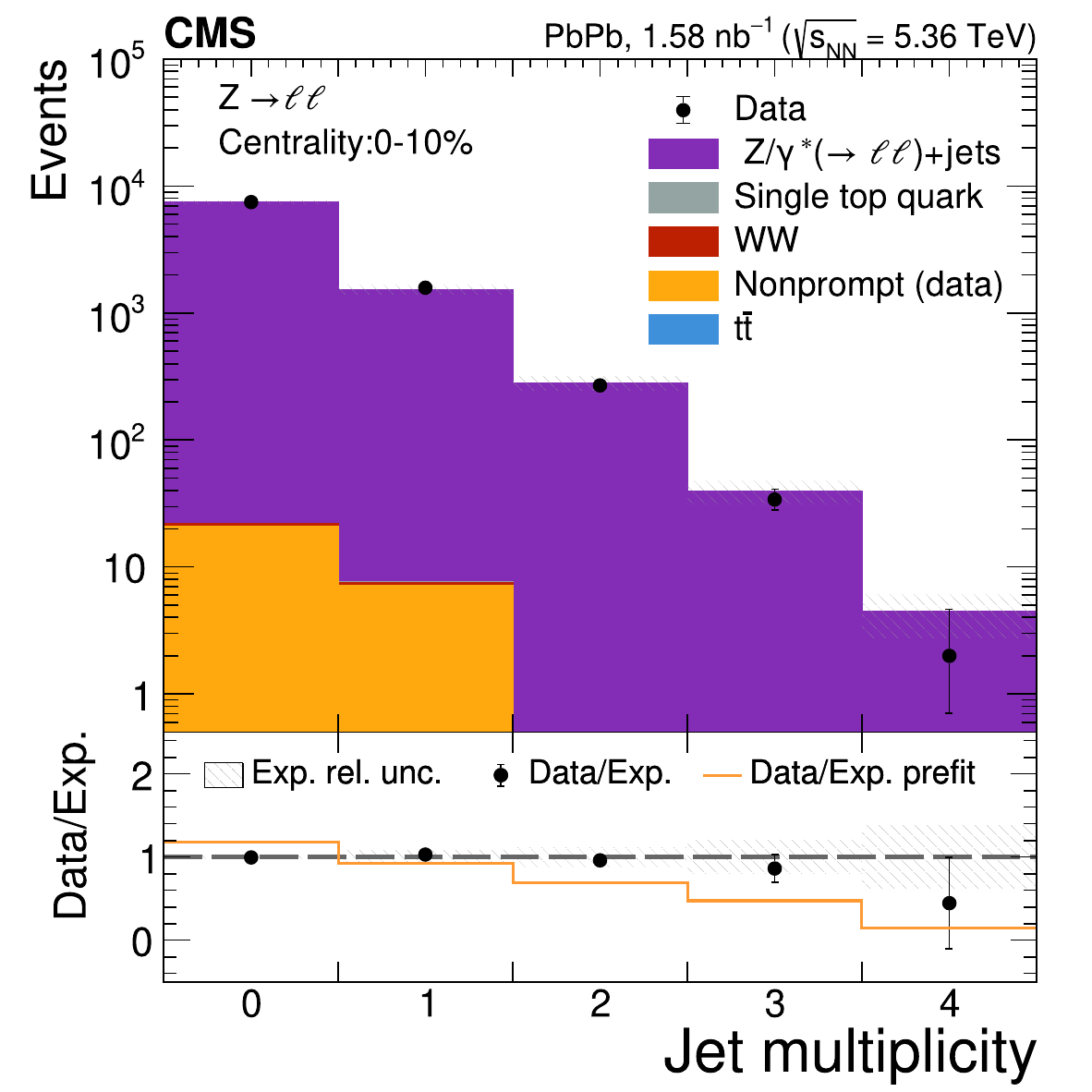}
\includegraphics[width=0.49\textwidth]{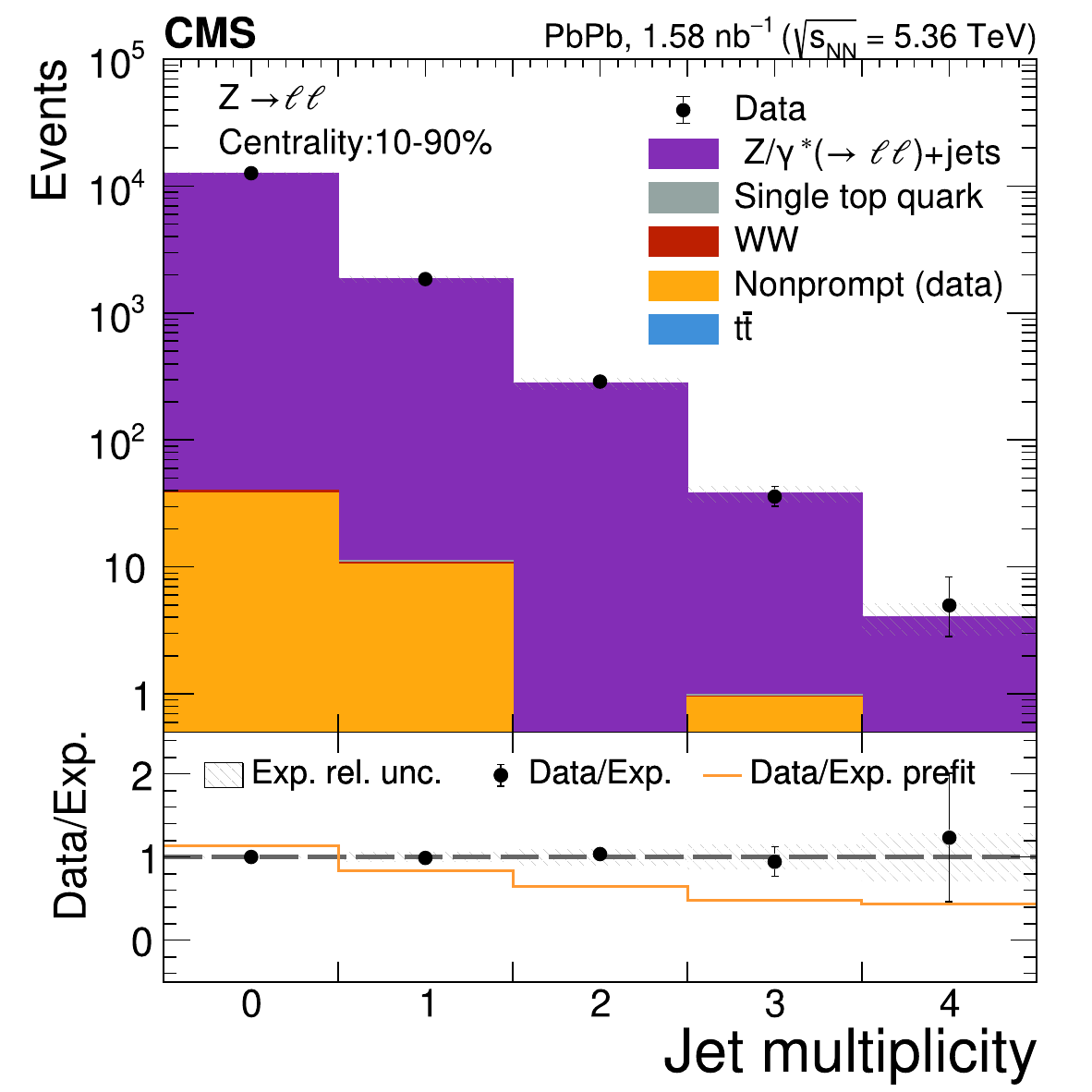}\\
\includegraphics[width=0.49\textwidth]{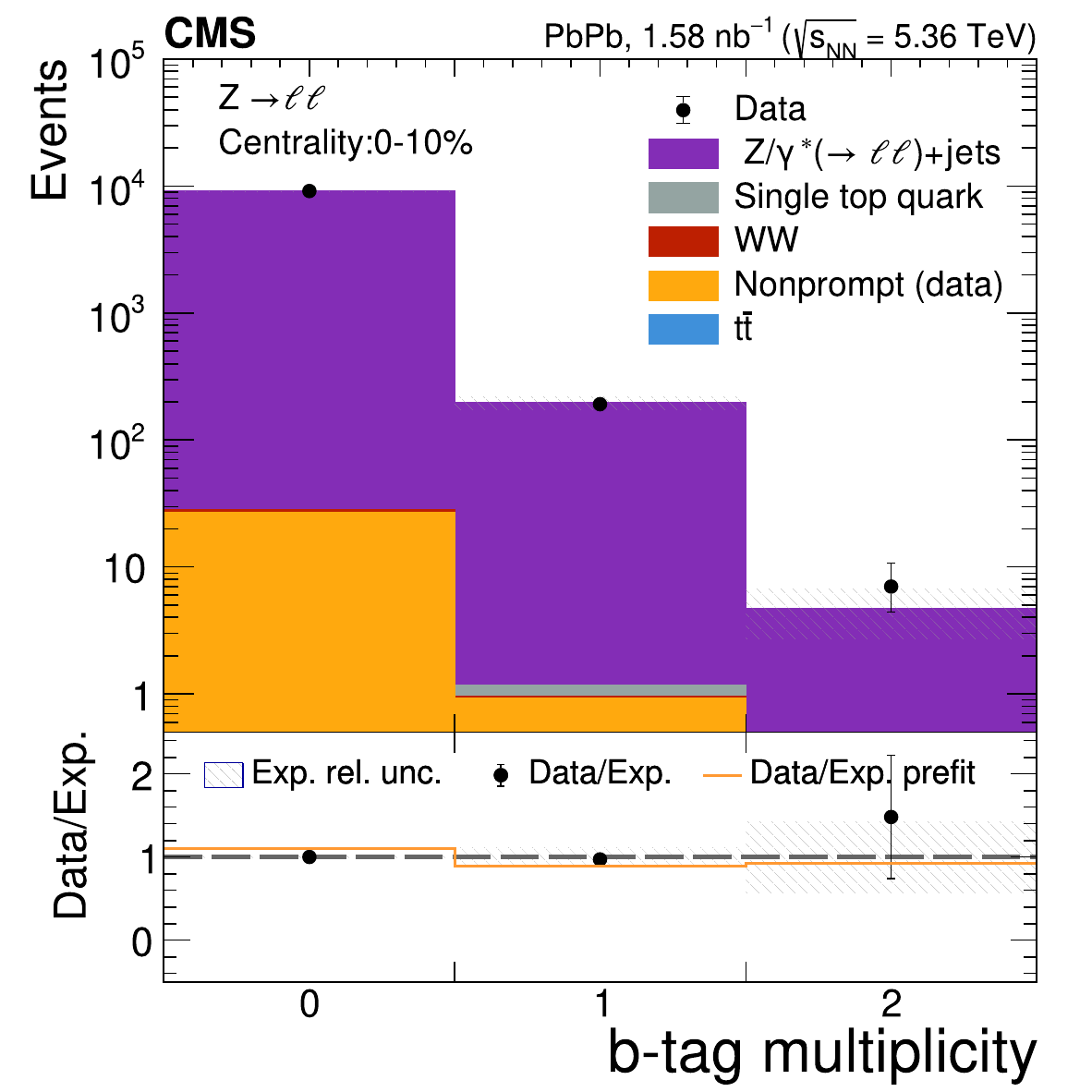}
\includegraphics[width=0.49\textwidth]{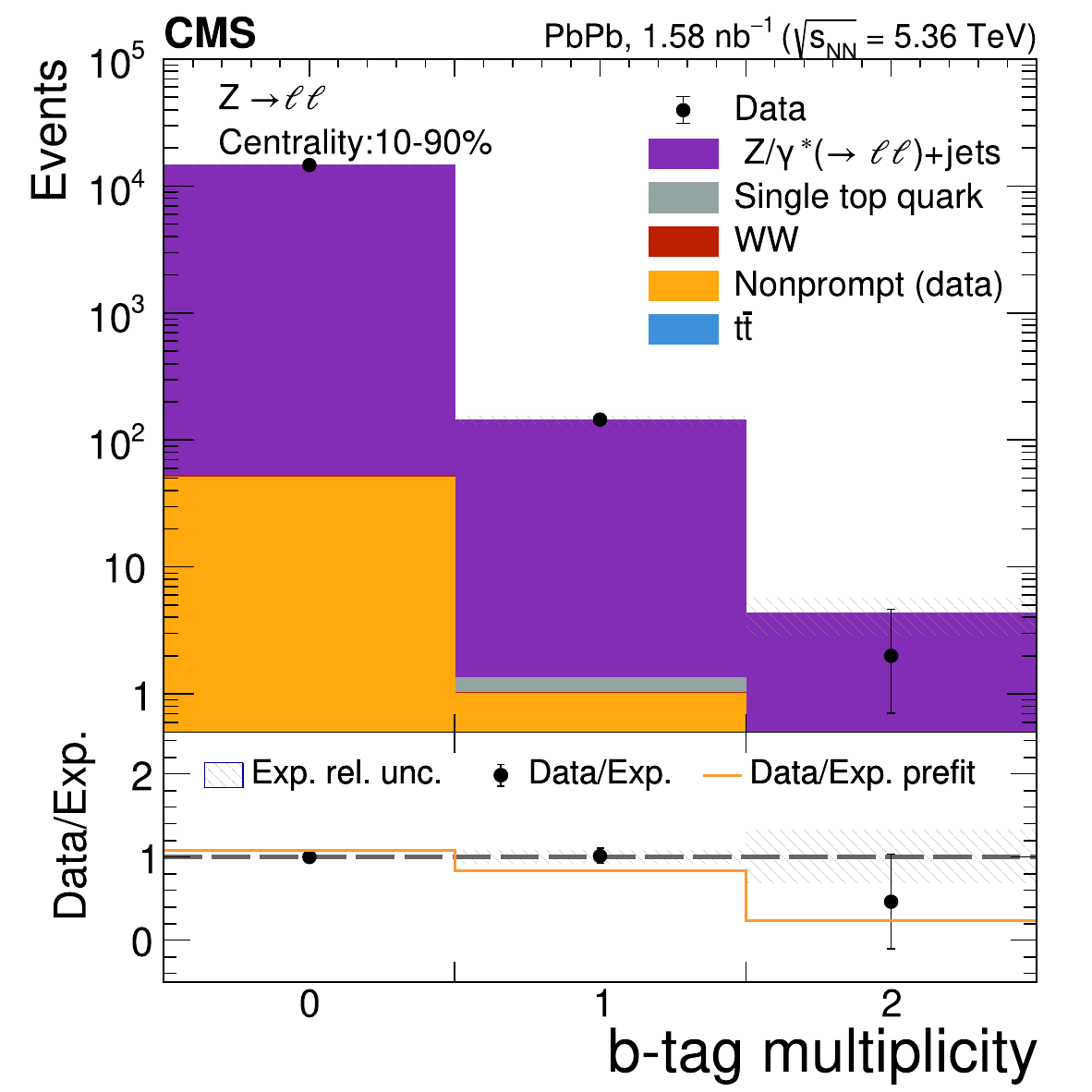}
\caption{Multiplicities of inclusive (upper) and \PQb-tagged (lower) jets in \PZ boson candidate events in the 0--10\% (left) and 10--90\% (right) centrality bins measured in the data (black markers) compared with the post-fit stacked model of the \ttbar (blue), nonprompt (light orange), \WW (red), single top quark (gray), and \DY (violet) contributions. The lower panels show the data-to-expectation ratios before (orange lines) and after (black markers) the fit. The hatched band represents the post-fit uncertainty of the model in each bin. The last bins include the overflow entries.}
\label{fig:njetfits}
\end{figure}

The adopted procedure is expected to provide a consistent description of the jets in data; any inconsistency would manifest as significant discrepancies in the number of reconstructed inclusive and/or bottom-quark jets. We therefore validate the approach by fitting the jet and \PQb jet multiplicities in \PZ boson events. The results of these fits are displayed in Fig.~\ref{fig:njetfits} for central (left) and semicentral (right) \PbPb\ collisions, and show a very good description of the inclusive jet (upper) and \PQb jet (lower) multiplicities.

\section{Background estimation}
\label{sec:backgrounds}

After the selection of dilepton pairs as described in Section~\ref{sec:dataandsel}, two main sources of background remain:
\DY production, which enters into the SF categories, and nonprompt leptons, which contribute mainly to the DF category.

The normalization of the DY background is left to float freely in the signal extraction fit described in Section~\ref{sec:statana}.
Prior to the fit, the counting of on-Z events shows good agreement between data and expectations within the uncertainties in the
integrated luminosity and theoretical cross section. The data/theory ratios amount to $1.08\pm0.04\stat$ for the \ee and $1.09\pm0.04\stat$ for the \mumu channels. As illustrated in Section~\ref{sec:statana}, good agreement is found between the measurements and the theoretical predictions in the BDT regions where \DY production is expected to dominate.

The second main source of background is nonprompt and misidentified leptons from multijet QCD events. Given the large cross section for the production of heavy-flavor jets, the probability for these processes to mimic a \ttbar dilepton event is not necessarily negligible for the working points used for identification and isolation of the lepton candidates. However, when pairing two leptons in the event, their properties are expected to be mostly uncorrelated with each other, particularly if the leptons had their origin in different partonic interactions.
This background is estimated based on an event-mixing technique which pairs leptons from different events as if they were produced in a single \PbPb collision.

The selection of events to be mixed is performed by identifying the nearest neighbors in the three-dimensional space defined by the
centrality, $\pt(\Pell\Pell)$, and $m(\Pell\Pell)$ bins. The distances between events are evaluated using a $k$-D tree algorithm,
which efficiently partitions the phase space in these variables and retrieves the closest matches~\cite{Bentley:1975bil,Brun:1997pa}.
For each event, four alternative dilepton candidates are constructed in this way, and the distribution of their kinematic
variables is used to obtain an alternative estimate of the nonprompt (combinatorial) background contribution.
The final estimate is an average of the four with the normalization assigned based on the number of SS events selected
in each channel. This technique significantly improves the statistical uncertainty associated with this background. Using simulated events, the accuracy of this method has been tested and ascribed an intrinsic uncertainty of 20--40\%, depending on the final state.

Finally, other less relevant backgrounds include single top quark and diboson (notably \WW) processes. These backgrounds are irreducible and estimated from the corresponding MC simulations described in Section~\ref{sec:MC} with the relevant data-to-MC corrections
applied to them.

\section{Signal extraction}
\label{sec:statana}

\begin{figure}[htbp]
\centering
\includegraphics[width=0.49\textwidth]{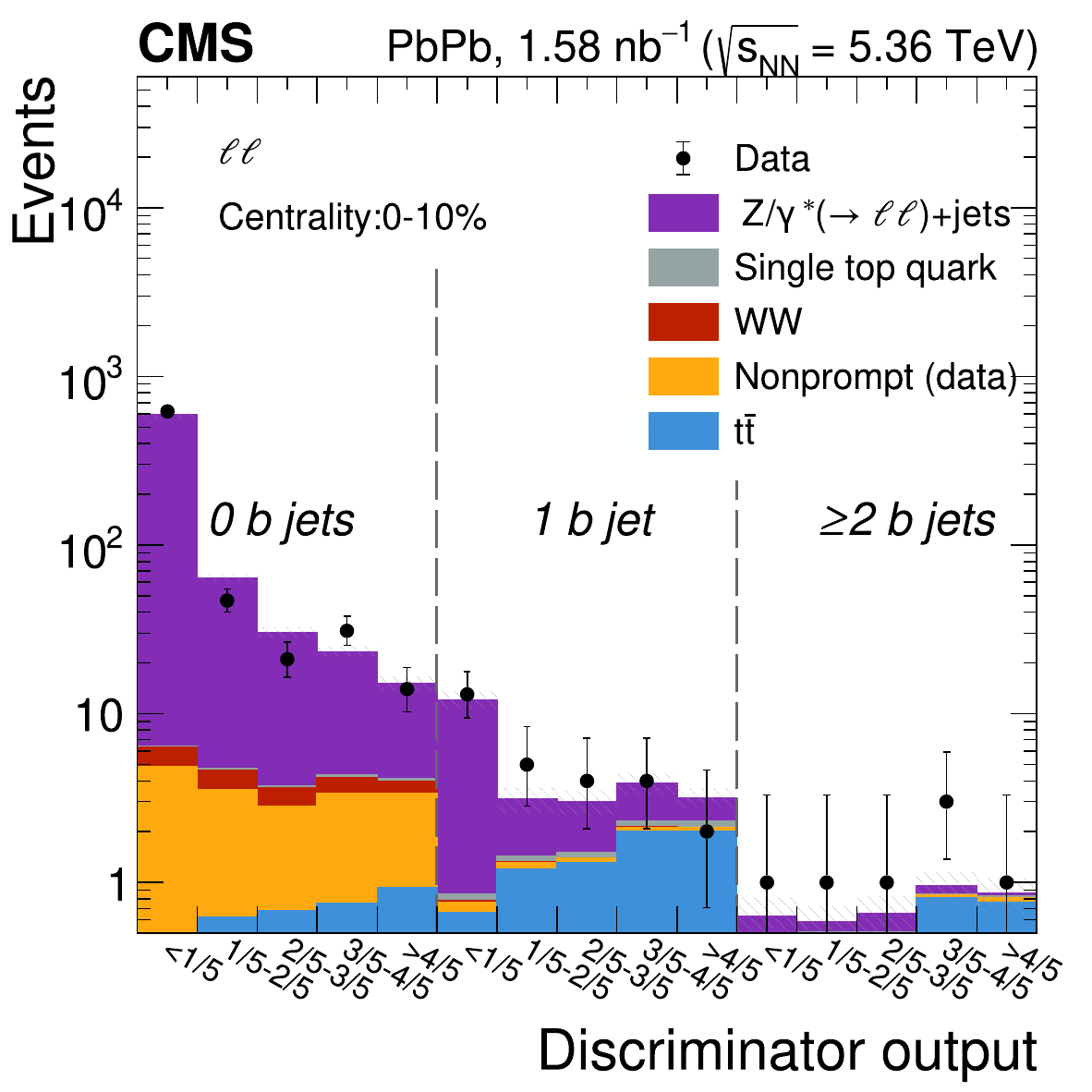}
\includegraphics[width=0.49\textwidth]{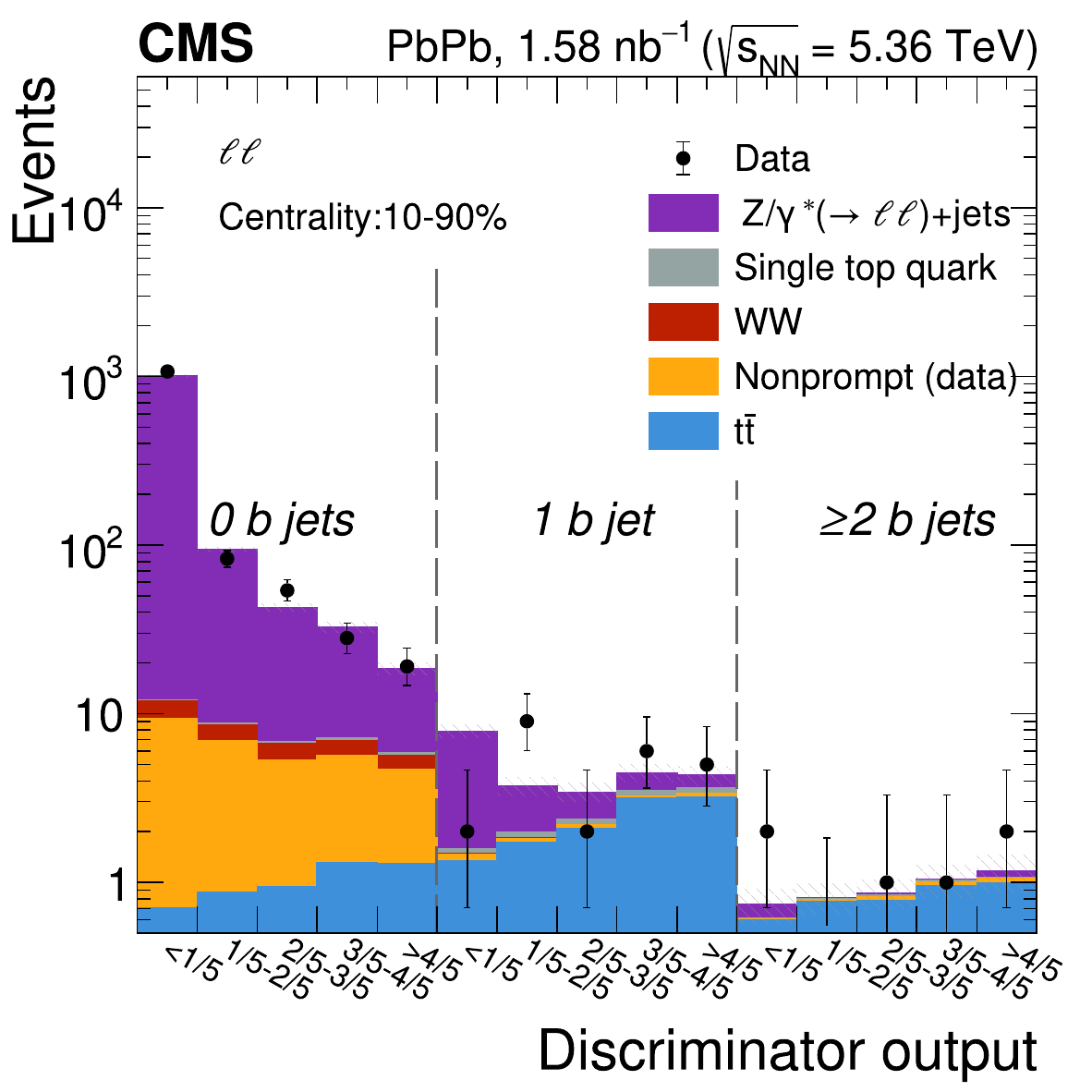}\\
\includegraphics[width=0.49\textwidth]{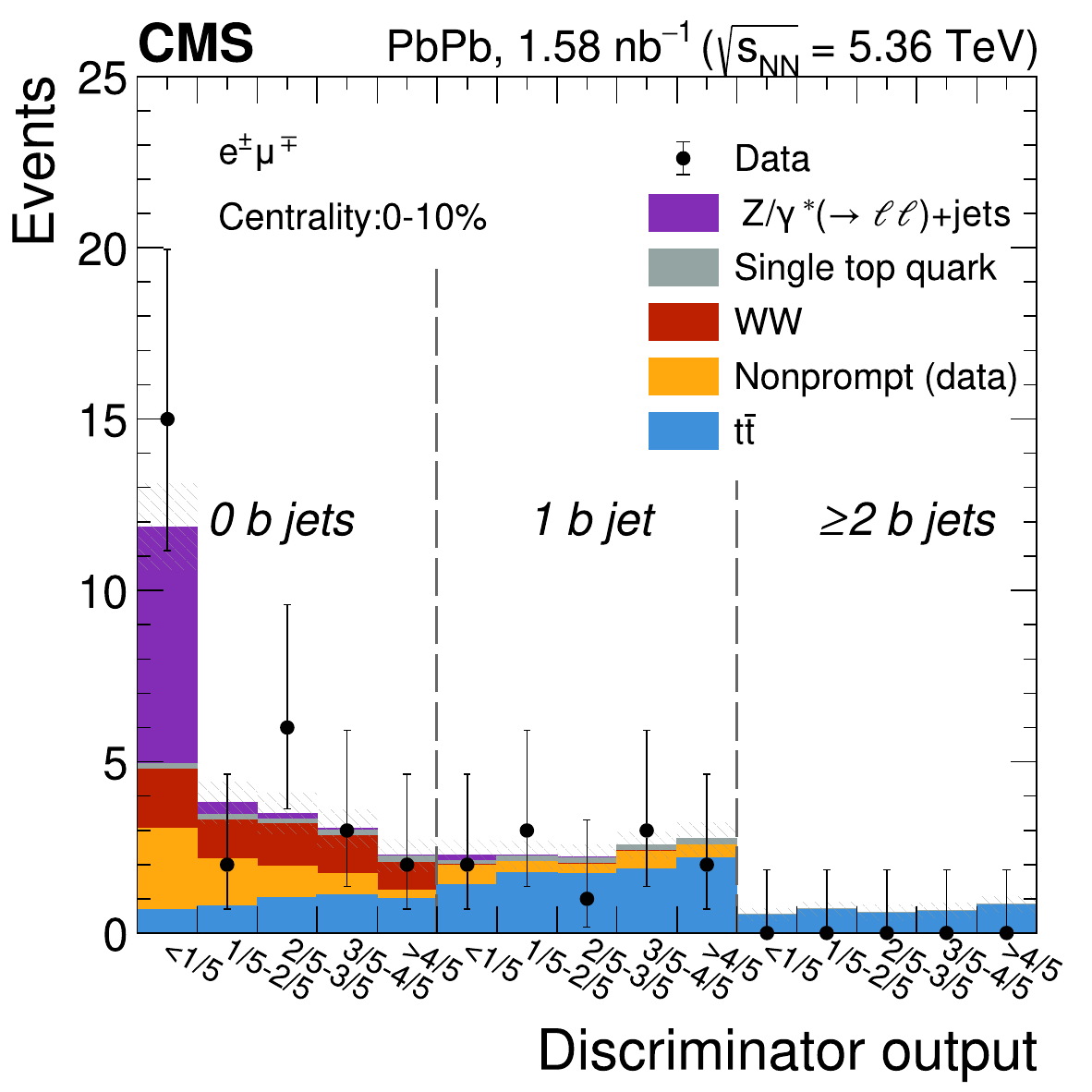}
\includegraphics[width=0.49\textwidth]{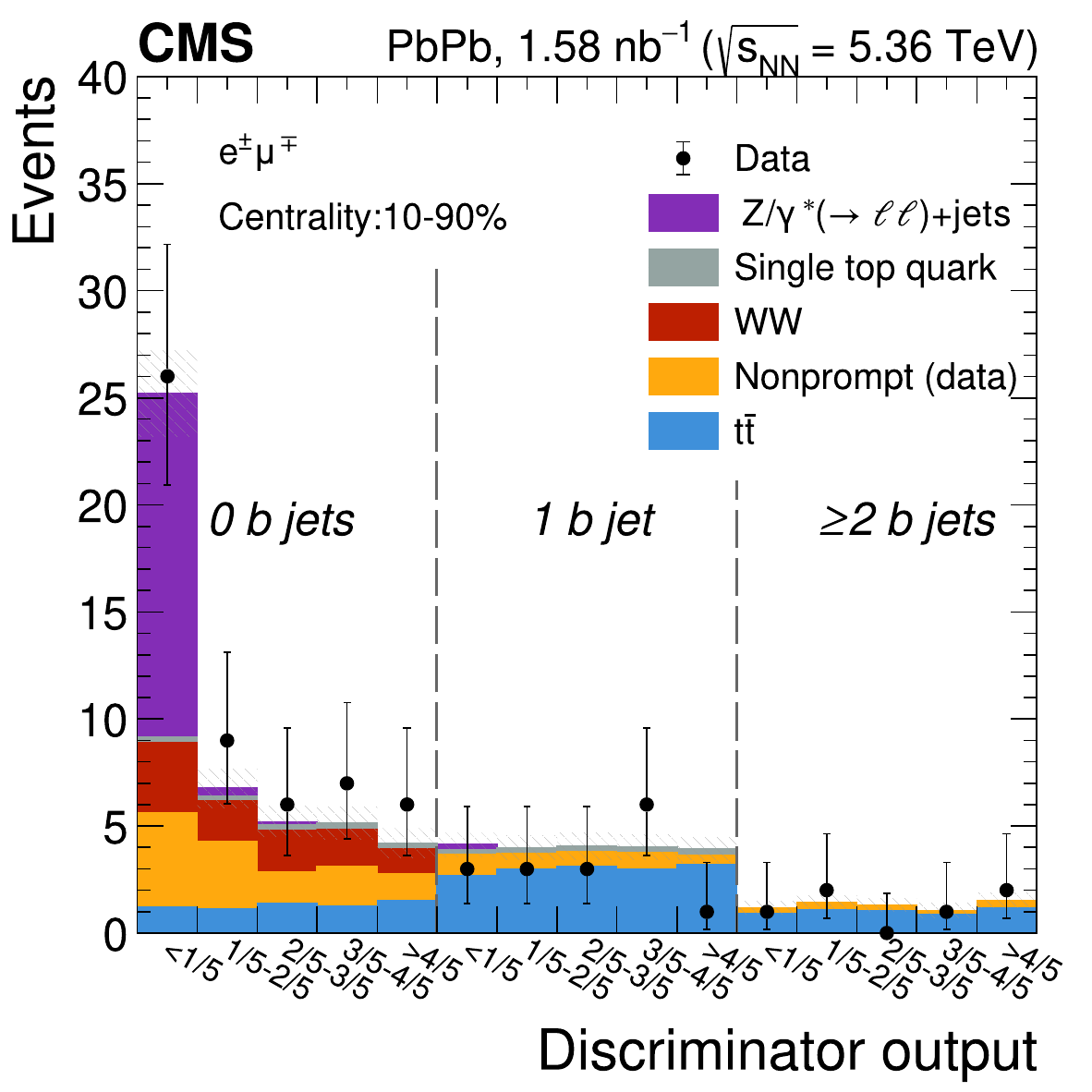}
\caption{Distributions of the final BDT discriminator after the fit for the same-flavor $\Pell\Pell$ (upper) and different-flavor \emu
(lower) channels. The discriminator output is shown for 0, 1, and $\geq2$ \PQb-tagged jet events. The left (right) figures correspond
to events reconstructed in the 0--10\% (10--90\%) centrality bin. The data (black markers) are compared with the stacked histograms
corresponding to the \ttbar signal (blue), and the nonprompt (light orange), \WW (red), single top quark (gray), and \DY (violet) background processes normalized to their post-fit expectations. The hatched gray bands show the total uncertainty in the predicted yields after the fit is performed.
\label{fig:bdt_ll_btag}}
\end{figure}

The \ttbar signal strength $\mu=\sigma_\text{meas}/\sigma_\text{th}$ (where $\sigma_\text{meas}$ and $\sigma_\text{th}$ are the measured and theoretical cross sections, respectively) is extracted from a fit to the BDT output of the \ee, \mumu (SF/off-Z/OS), and \emu (DF/OS) channels, analyzed in events with zero, one, or at least two \PQb-tagged jets. Two additional categories ($\PZ\to\ee$ and $\PZ\to\mumu$ in the on-Z/OS class) are added to the fit to control the normalization of the \DY background. In each category, the fit is performed to the BDT output in events with zero, one, or at least two \PQb-tagged jets. The impact of the systematic uncertainties on the signal and background expectations, detailed below, is modeled through a set of dedicated nuisance parameters~\cite{Conway:2011in}. The fitted BDT distributions are displayed in Fig.~\ref{fig:bdt_ll_btag}.
In addition to the signal strength and the corresponding breakdown of statistical and systematic uncertainties, the significance of the \ttbar process is also extracted. The statistical analysis is performed using the \textsc{Combine} tool~\cite{CMS:2024onh} based on the \textsc{RooFit}~\cite{Verkerke:2003ir} and \textsc{RooStats}~\cite{Moneta:2010pm} frameworks. The best-fit signal strength and its approximate 68\% confidence level (\CL) interval are extracted following the procedure outlined in Section~3.2 of~\cite{Khachatryan:2014jba}, while the significance ($Z$ score) is evaluated using an asymptotic approximation and validated with pseudo-experiments~\cite{pvalue}. The experimental uncertainties can affect the modeling of both signal and backgrounds in the fit.
They can be broadly grouped into those associated with the selection and calibration of physics objects,
and those arising from the background estimation, as discussed below.

The efficiency of the trigger and lepton selections is corrected for the differences between data and simulation using scale factors
derived from events in the $\PZ\to\Qlp\Qlm$ CRs. The uncertainties associated with these scale factors are propagated to
the analysis by varying the event weights used in the simulation, resulting in different predictions for the cross sections for the
signal, single top, and diboson processes. The difference between these alternative predictions and the nominal one is parameterized as a function of several nuisance parameters depending on the type of lepton and the associated efficiency (trigger, identification, or
isolation).

The impact of jet reconstruction inefficiencies in regions where CMS subdetectors suffer from aging or malfunctioning components is evaluated. These areas are mapped in the $\eta$--$\phi$ plane, and simulated jets with hadronic activity inside them are removed from the event. The resulting differences in the BDT shapes, relative to the nominal simulation, are assigned as a systematic uncertainty.
Additional sources of jet selection inefficiency may arise from the \textsc{UParT} application, as discussed in Section~\ref{sec:dataandsel}. To assess this effect, the data-to-simulation scale factors associated with the \textsc{UParT} efficiency are independently varied for UE jets and for jets matched to the hard process in the simulation. The corresponding variations in the BDT distributions are taken as a systematic uncertainty.

The impact of the jet energy scale uncertainty on the final observables is broken down into four main contributions: parton type, hadronization, process dependence, and uncertainty in the residual scale. The parton type uncertainty is estimated from simulation based on the different responses of \PQb, and light-quark or gluon jets with respect to the mixed-flavor response. The hadronization uncertainty is evaluated from the difference in detector response obtained when using \PYTHIA versus \HERWIG7~\cite{Bahr:2008pv,Bellm:2015jjp} as parton shower and hadronization models. The process-dependent uncertainties result from the fact that \textsc{UParT} has been trained using \ttbar events, and it may thereby exhibit residual response differences when applied to other final states with different jet kinematics. This effect is evaluated from the jet response in the \Zjets MC sample. Finally, the residual scale uncertainty is subdivided into two sources: jet quenching and extrapolation to the vacuum or peripheral collision regime. Both are estimated from the average $x_\mathrm{j}$ ratio measured in data over the MC prediction (Fig.~\ref{fig:rjetfits_2}). The deviation from unity when extrapolating the relative response to peripheral events is considered as a ${\approx}5\%$ systematic uncertainty. The systematic uncertainty associated with jet quenching is evaluated by comparing the final observables with and without applying the parameterized energy loss shown in Fig.~\ref{fig:rjetfits_2} on a jet-by-jet basis in the simulation. This procedure leads to a substantial loss of efficiency for reconstructing jets with $\pt>30\GeV$, thereby affecting the jet multiplicity in the \PQb-tagged analysis. Finally, the impact of \PQb tagging on the predicted distributions shown in Fig.~\ref{fig:bdt_ll_btag} is assessed by varying the \PQb tagging working point, \ie using loose or tight instead of intermediate, such that the signal efficiency changes by ${\pm}10\%$. These variations correspond to an expected change of the misidentification rate by more than a factor of two~\cite{CMS-DP-2024-088}.

Beyond the uncertainties affecting the reconstruction of physics objects, the normalization of the event yields is impacted by a 6.4\%
uncertainty in the total \PbPb integrated luminosity obtained from a preliminary analysis of van~der~Meer scan data collected during the \PbPb run, following the same methodology of Ref.~\cite{CMS:2025rzq}.

All simulated signal and background samples are affected by the experimental uncertainties described above. In the signal extraction fit, these uncertainties are treated as fully correlated between signal and background processes, with the exception of the \PQb tagging uncertainties. The latter are kept uncorrelated to reflect the different parton type and kinematics between the on-\PZ CR and the \ttbar signal region. Theoretical uncertainties from missing higher-order corrections in the ME predictions are evaluated by varying the $\mu_\text{F}$ and $\mu_\text{R}$ scales within factors of two (excluding variations in opposite directions). The nuisance parameters associated with these scale uncertainties are treated as uncorrelated between the background and signal samples. The normalization of the dominant DY background is left unconstrained in the fit, but an additional model uncertainty is introduced from the difference between the shape of the \POWHEG and \MGvATNLO predictions. For the nonprompt-lepton background, uncertainties arise from its normalization (20--40\%), and from differences in shape between the event-mixing method and the SS and nonisolated sidebands. An uncertainty of 30\% in the absolute normalization of other backgrounds (single top and \PW\PW) has been included in the fit.
Statistical uncertainties in the simulated background differential cross sections are also incorporated via a Barlow--Beeston ``lite'' treatment~\cite{Barlow:1993dm}.

Finally, the signal MC samples include two classes of modeling uncertainties. First, differences between the signal BDT shapes predicted by \POWHEG and \MGvATNLO are included as a ME generator uncertainty. Although this variation is nominally one-sided, a symmetrized uncertainty is used in the statistical analysis. Additional modeling uncertainties are introduced by varying the top quark mass by ${\pm}1\GeV$ (implemented through generator-level reweighting using relativistic Breit--Wigner ratios) and by modifying the top quark \pt spectrum through an empirical reweighting that reproduces the softer \pt distribution observed in the $\Pp\Pp$ data~\cite{CMS:2024ybg}. Second, uncertainties from the parton densities are evaluated by reweighting the \ttbar events using all eigenvectors of the EPPS21\_nlo set---a total of 106 variations---and propagating the resulting changes in the BDT output. The final nPDF plus $\alpS$ uncertainties are determined following the PDF4LHC prescription~\cite{Butterworth:2015oua}.

\section{Results}
\label{sec:results}

The distributions of the dilepton BDT discriminators in the different \PQb-tagged jet multiplicity categories are shown in Fig.~\ref{fig:bdt_ll_btag}. In all plots, the event yields of the processes are normalized to the values determined by the fit. Fair agreement between data and post-fit expectations is obtained in the regions dominated by the backgrounds. In high-purity signal regions, the statistical fluctuations remain sizeable.
The BDT discriminator exhibits strong separating power between the signal and the various background processes throughout all spectra.
The inclusion of the \PQb tag multiplicity yields additional discriminating power, improving separation even in low-purity regions.

\begin{figure}[htbp]
\centering
\includegraphics[width=0.8\textwidth]{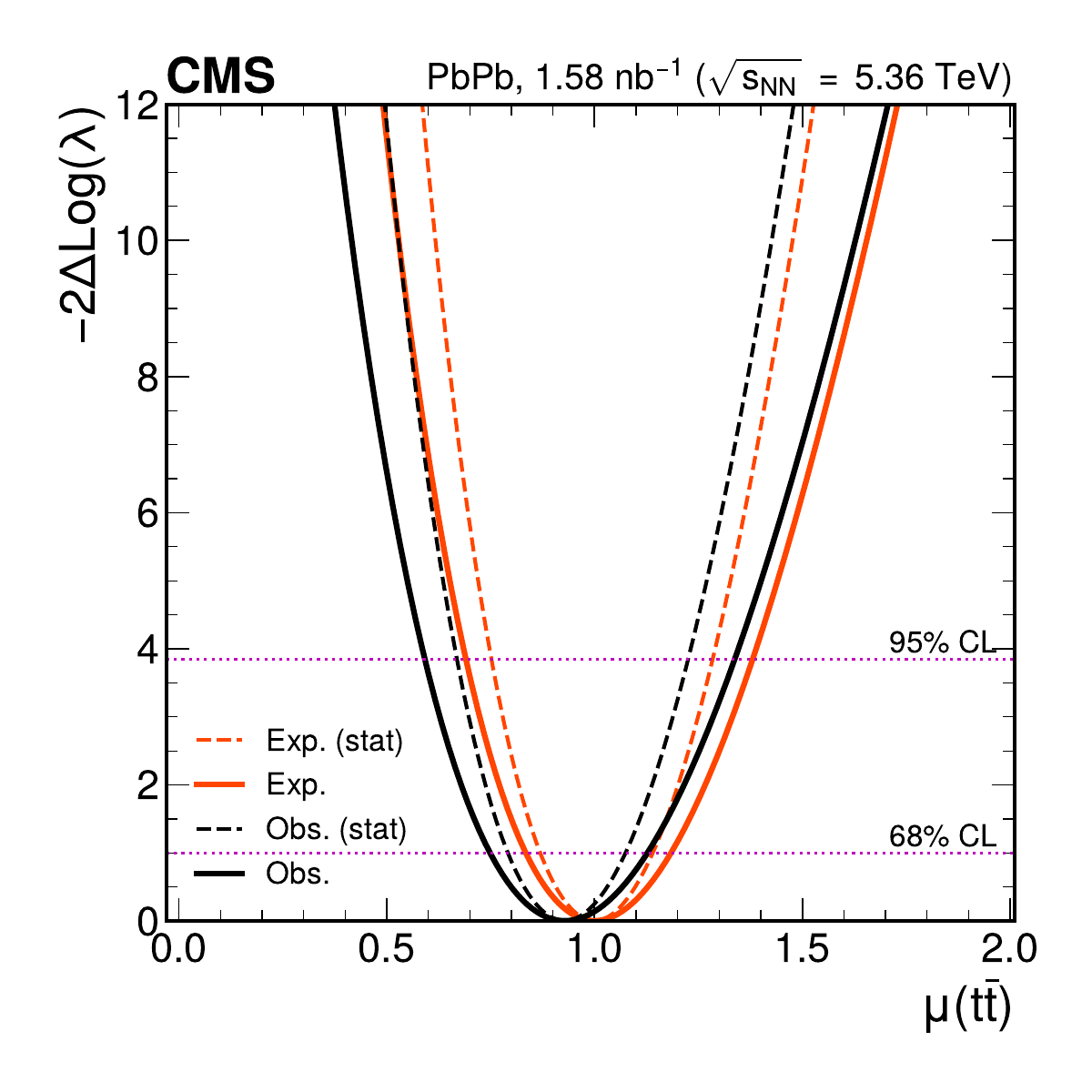}
\caption{Scan of the profile likelihood as a function of the \ttbar signal strength. The expected (orange) and observed (black) scans are displayed with (solid curve) and without (dashed curve) accounting for systematic uncertainties. The horizontal dashed lines indicate the likelihood values used to extract the 68\% and 95\% \CL intervals on the parameter of interest.
\label{fig:nll_scan}}
\end{figure}

Through the data fitting procedure described in Section~\ref{sec:statana}, the \ttbar yield is extracted with a signal strength of
$\mu(\ttbar)=0.93^{+0.15}_{-0.14}\stat^{+0.14}_{-0.12}\syst$, corresponding to an observed significance exceeding five standard deviations relative to the background-only hypothesis~\cite{Cowan:2010js}. The likelihood scan is shown in Fig.~\ref{fig:nll_scan}. The statistical
uncertainty remains slightly larger than the systematic component. Systematic uncertainties, particularly those associated with object efficiencies and background modeling, are expected to decrease further as larger data samples become available in the future.

\begin{figure}[htbp]
\centering
\includegraphics[width=0.8\textwidth]{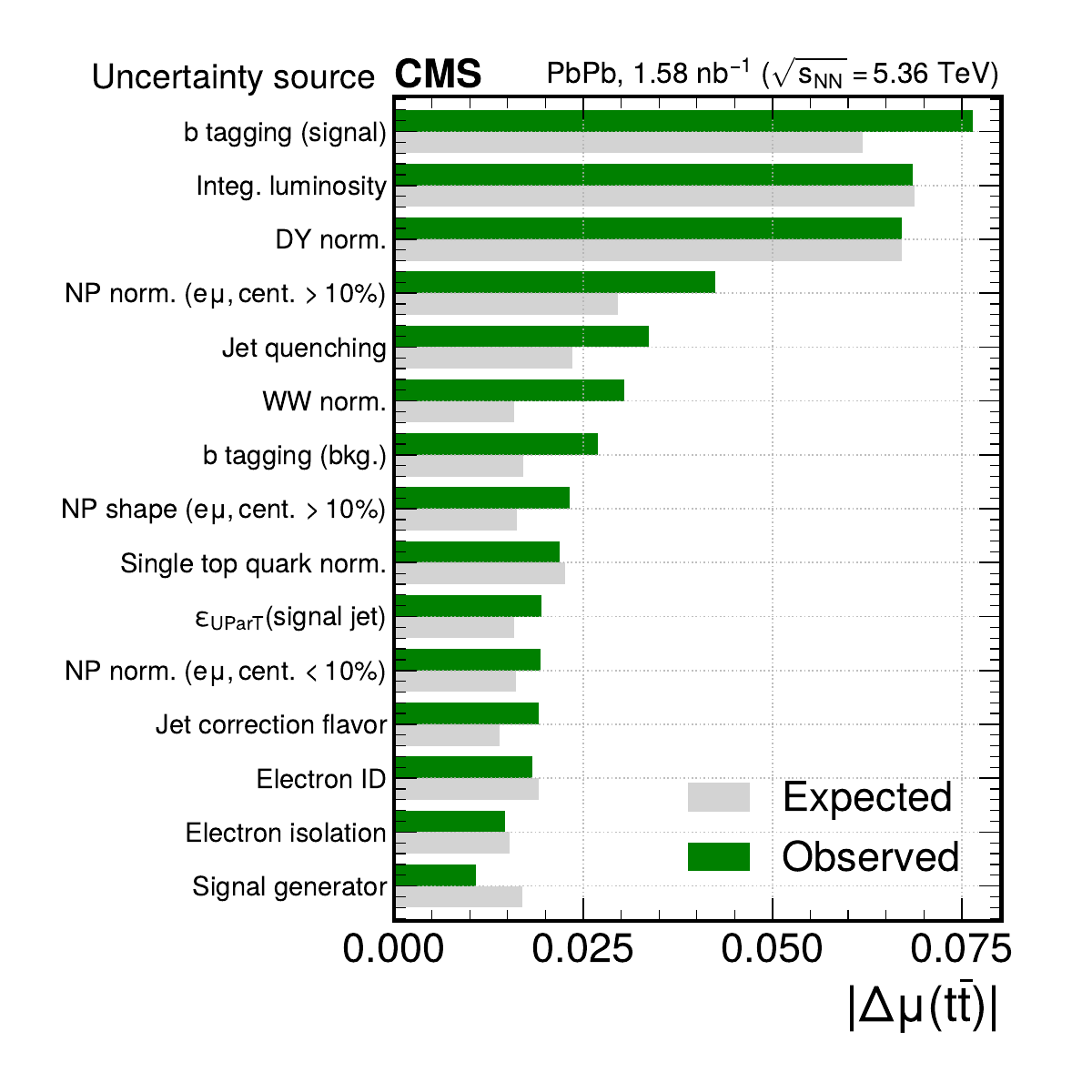}
\caption{Impact of systematic uncertainties on the fitted \ttbar signal strength parameter. The systematic uncertainties are listed in
decreasing order of their impact on the signal strength.}
\label{fig:impacts}
\end{figure}

The main systematic uncertainties arise from the \PQb tagging, integrated luminosity, and the normalization in the DY and nonprompt-lepton backgrounds.
The impact of each source of systematic uncertainty in the extracted signal strength is quantified by evaluating the shift induced when the corresponding nuisance parameter is varied up or down by its post-fit uncertainty~\cite{CMS:2024onh}. Figure~\ref{fig:impacts} shows the resulting impact magnitudes for the dominant sources of uncertainty.
The $p$-value associated with the goodness-of-fit is estimated to be 10\%. Several cross-checks have been performed to assess the
robustness of the fit, including separate analyses of the different final states, the omission of \PQb tagging requirements, the
exclusion of the \PZ boson CRs from the fit, and measurements performed in exclusive centrality categories.
In all cases, the results are found to be compatible with the signal strength derived in the default analysis.

{\tolerance=800 From the signal strength, a cross section $\sigma_{\ttbar}=\sigmaTTbarobs\mub$ is derived, which is in agreement with the pQCD predictions at NNLO\,$+$\,NNLL accuracy. The DY cross section, measured in a phase space where $m(\Pell\Pell)>10\GeV$, is found to be $\sigma_\text{DY}=\sigmaDYobs\mub$, with uncertainties dominated by the \PbPb integrated luminosity, also in agreement with the theoretical expectations ($\sigma^\text{th}_\text{DY}=370\mub$ for the central NNLO value obtained with EPPS21\_nlo).
Partial reduction of some of the main systematic uncertainties, such as the integrated luminosity and lepton efficiencies, can be achieved by measuring the ratio of the \ttbar over the DY cross sections. In addition, given that \ttbar is
preferentially produced via gluon-gluon fusion, whereas DY pairs through quark-antiquark annihilation, this ratio directly
probes the relative gluon and quark content of the nuclear PDFs. The ratio of the two cross sections is evaluated in the full
phase space for the \ttbar process and for pair masses $m(\Pell\Pell)>10\GeV$ for the DY process. The extrapolation of the visible
signal strength of each process to the full phase space results in an intrinsic uncertainty of $^{+2.8\%}_{-3.1\%}$, estimated using NLO pQCD predictions.
The measured ratio amounts to $\RttDY=\RTTbarDYobs$
in agreement with the theoretical ratio at NNLO accuracy, $\RttDY^\text{th}=0.0108^{+0.0026}_{-0.0023}$ (nPDF+scales), obtained with the
EPPS21\_nlo nPDFs using \TOPpp and \textsc{dyturbo}~\cite{Camarda:2019zyx} for the numerator and denominator, respectively.\par}

\begin{figure}[htbp]
\centering
\includegraphics[width=0.45\textwidth]{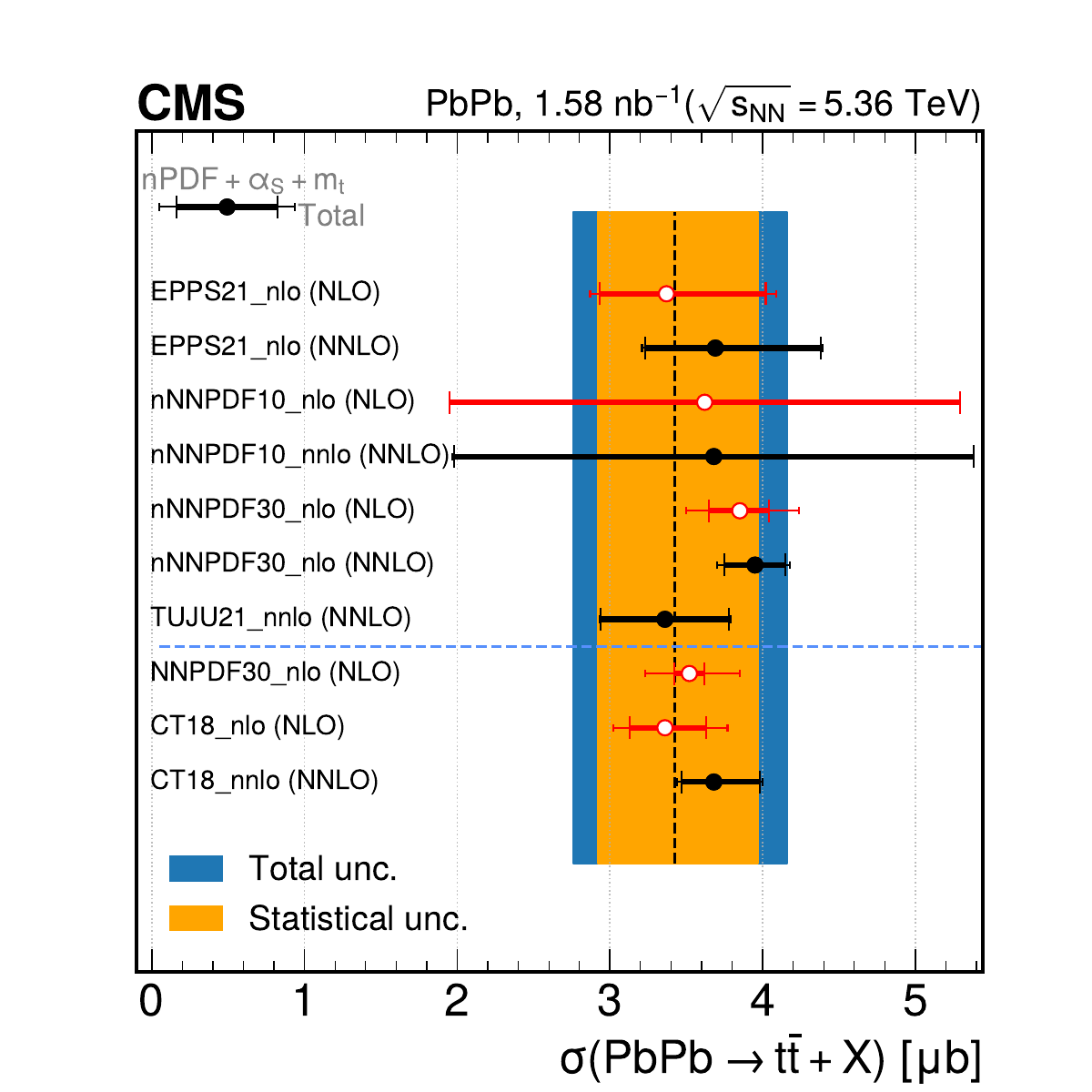}\\
\includegraphics[width=0.45\textwidth]{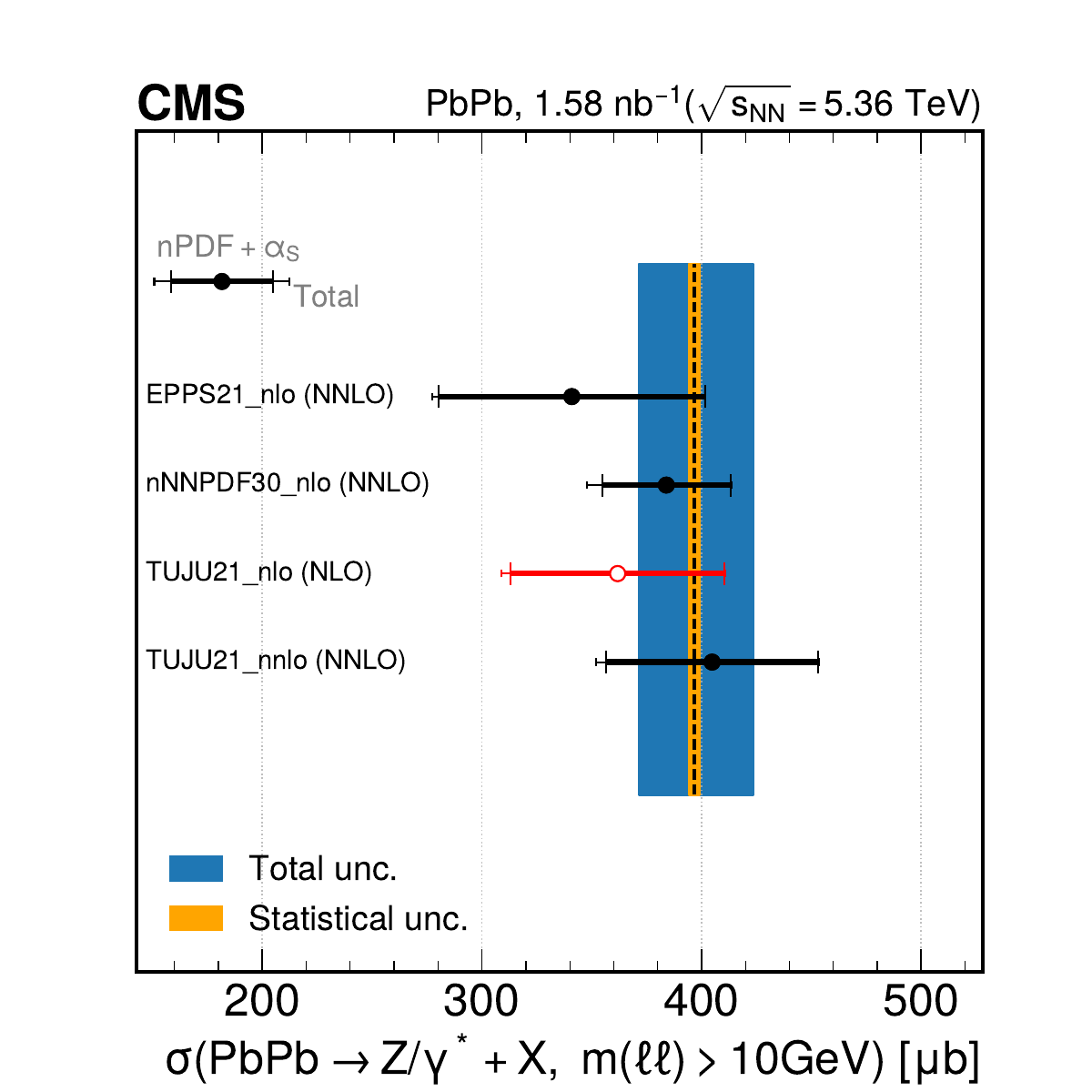}\\
\includegraphics[width=0.45\textwidth]{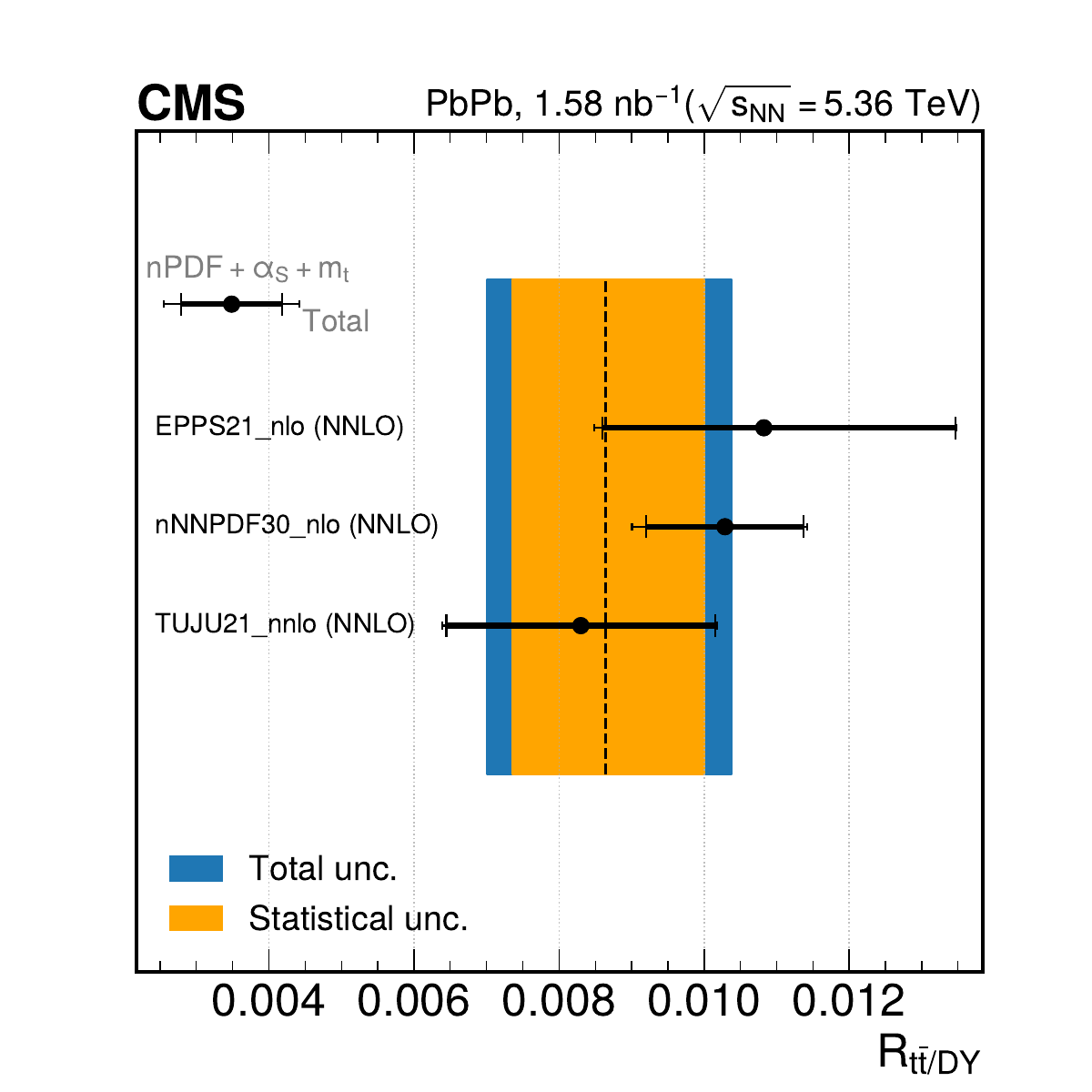}
\caption{Experimental measurements (colored bands) of the \ttbar cross section (upper), DY cross section (middle), and $\RttDY$ ratio (lower) in inclusive \PbPb collisions at $\sqrtsNN=5.36\TeV$ compared with the corresponding NLO (red bars) or NNLO\,$+$\,NNLL (black bars) theoretical predictions obtained with the EPPS21\_nlo, nNNPDF10\_(n)nlo, nNNPDF30\_nlo, and TUJU21\_(n)nlo nPDF sets. The error bars represent the theoretical uncertainties, and the blue (orange) area shows the total (statistical) uncertainties of the measurements. In the upper plot a dashed line separates the theory predictions obtained with nuclear PDFs (above the line) from those with the proton PDFs (below the line).}
\label{fig:summary}
\end{figure}

{\tolerance=800 Figure~\ref{fig:summary} compares the experimental \ttbar (upper) and DY (middle) cross section as well as $\RttDY$ ratio (lower) with the theoretical predictions obtained at NLO or NNLO pQCD accuracy (as indicated in parentheses in the legends) using different PDFs: EPPS21\_nlo~\cite{Eskola:2021nhw}, nNNPDF10\_(n)nlo~\cite{AbdulKhalek:2019mzd}, nNNPDF30\_nlo~\cite{AbdulKhalek:2022fyi}, and TUJU21\_(n)nlo~\cite{Helenius:2021tof} nuclear sets, as well as NNPDF30\_nlo~\cite{NNPDF:2014otw}, and CT18\_(n)nlo proton PDF sets. The statistical and total experimental uncertainties are shown as orange and blue bands in the plots, respectively. The black error bars denote the total theoretical uncertainty; the inner bars indicate the contribution from nPDF plus parametric ($\alpS$ and $m_{\PQt}$) uncertainties. Overall, the measurements are in good agreement with the theoretical predictions within uncertainties for all observables. Considering the theoretical predictions for the \ttbar production cross sections, several features are noteworthy. First, for any given (n)PDF set, the NNLO predictions are larger than their NLO counterparts. Second, again for the same (n)PDF input, the NNLO calculations exhibit reduced uncertainties---particularly those associated with theoretical scale variations---relative to the NLO results. Third, the cross sections obtained using proton PDFs are a few percent smaller than those derived with nuclear PDFs, reflecting a mild antishadowing effect in the nuclear parton densities~\cite{dEnterria:2015mgr}. Finally, the substantially larger uncertainties of nNNPDF1.0 compared with the more recent nNNPDF3.0 set are expected, since the former relies solely on legacy fixed-target nuclear data, whereas the latter incorporates a broad range of modern LHC measurements.
Although the current experimental precision of the \ttbar cross section does not yet allow discrimination among predictions based on different nPDF sets, it is worth noting that the corresponding theoretical uncertainties are of comparable magnitude to the combined measurement presented here. In contrast, for the DY cross section and the $\RttDY$ observable, the uncertainties on the theoretical predictions are, in several cases, larger than the experimental ones. These measurements are therefore expected to provide new and meaningful constraints in future global nPDF fits, particularly by probing kinematic regions that have not been accessible in previous analyses. \par}

\begin{figure}[htbp]
\centering
\includegraphics[width=0.45\textwidth]{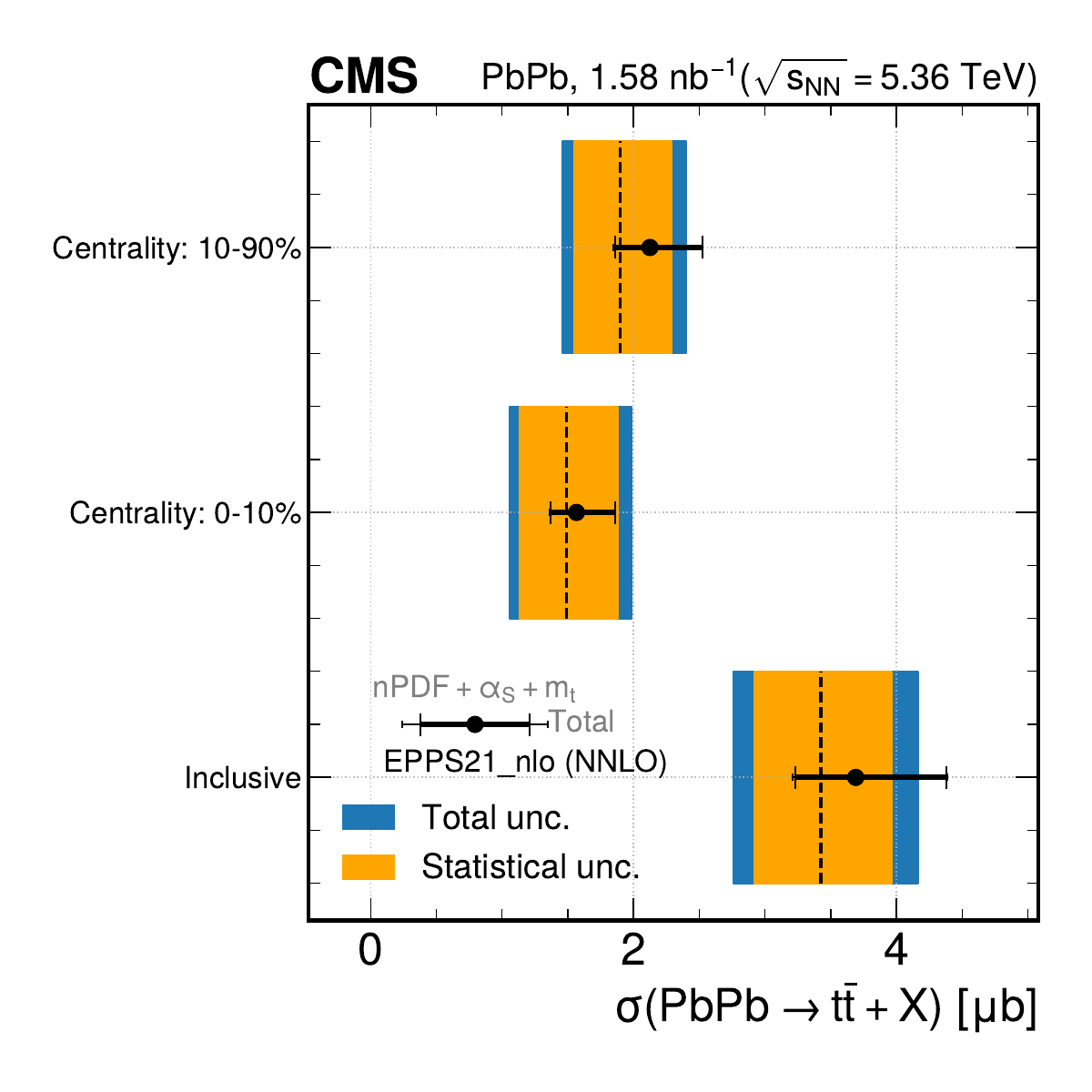}\\
\includegraphics[width=0.45\textwidth]{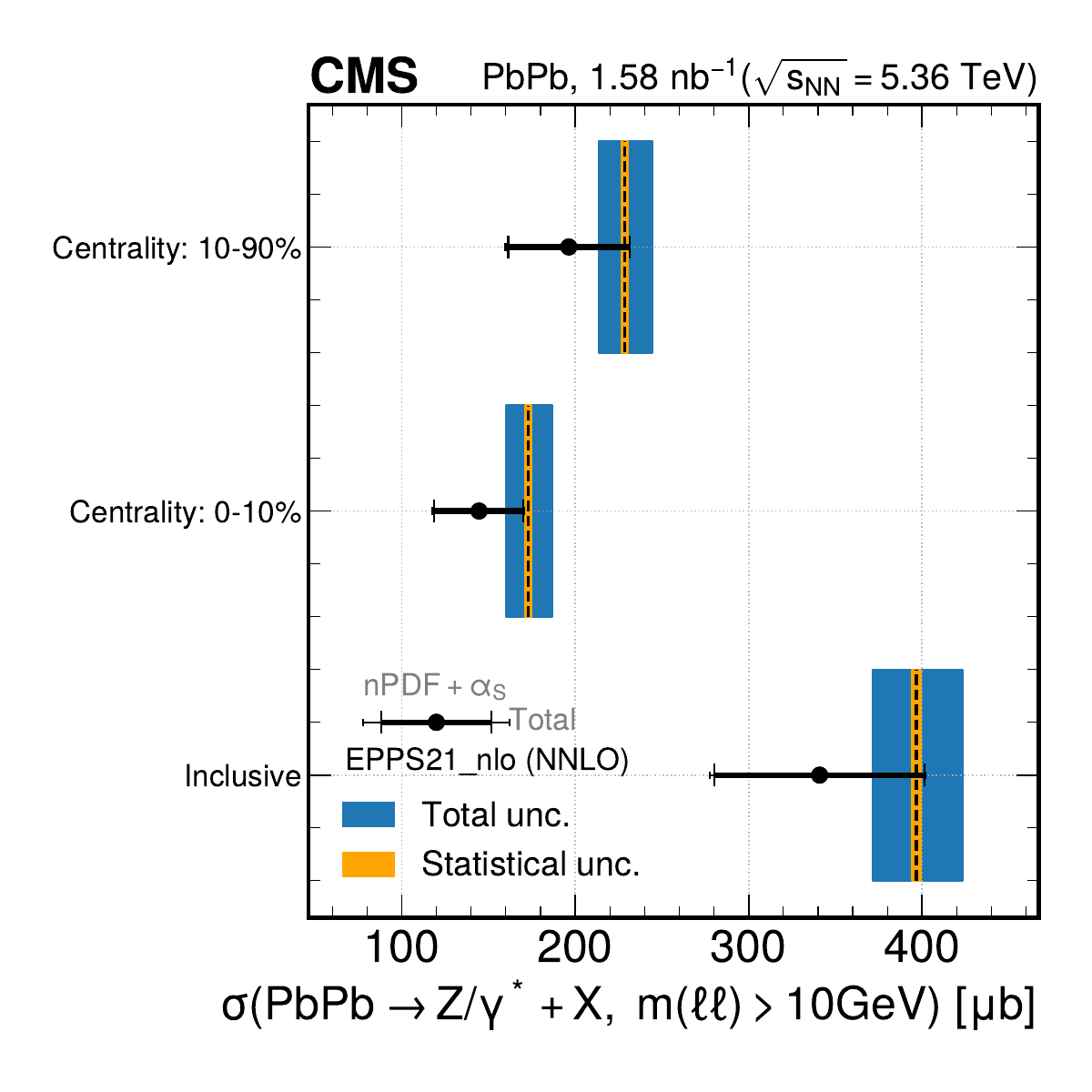}\\
\includegraphics[width=0.45\textwidth]{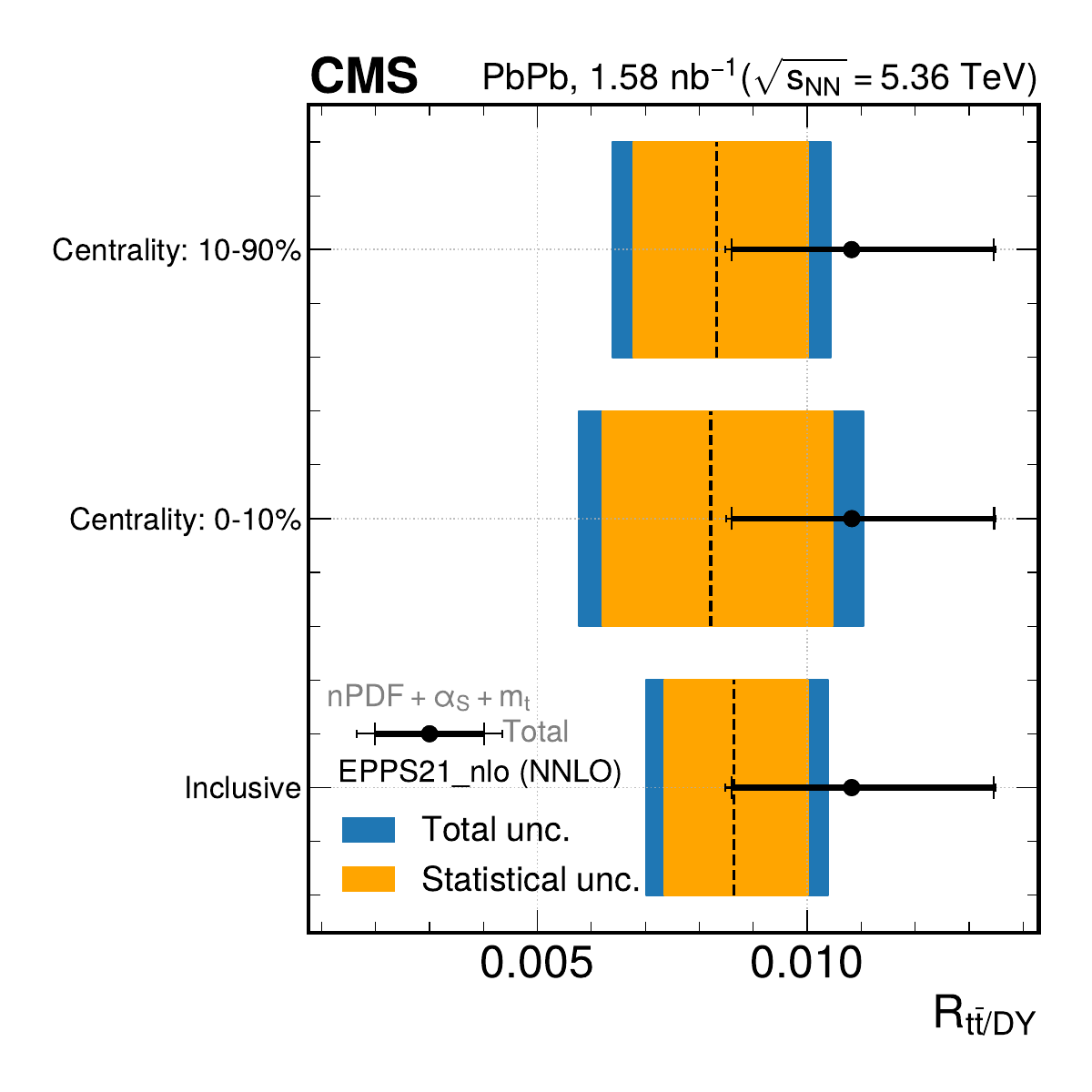}
\caption{Experimental measurements (colored bands) of the \ttbar cross section (upper), DY cross section (middle), and $\RttDY$ ratio (lower) in semicentral (10--90\%), central (0--10\%), and inclusive \PbPb collisions at $\sqrtsNN=5.36\TeV$ compared with the corresponding NNLO\,$+$\,NNLL theoretical predictions (black markers with error bars) obtained with the EPPS21\_nlo~\cite{Eskola:2021nhw} nPDF set. The blue (orange) bands indicate the total (statistical) uncertainties of the measured observables, whereas the error bars represent the uncertainties in the theory calculations.}
\label{fig:diff_summary}
\end{figure}

In addition, the \ttbar and DY cross sections, as well as their ratio, are measured separately for 0--10\% central and 10--90\% semicentral \PbPb collisions. Table~\ref{tab:diff_summary} summarizes all the experimental results obtained. Figure~\ref{fig:diff_summary} shows the different observables measured in the three \PbPb centrality selections (10--90\%, 0--10\% and inclusive) compared with the corresponding theoretical predictions obtained at NNLO\,$+$\,NNLL pQCD accuracy with the EPPS21\_nlo set. In all cases, the measurements are consistent with the pQCD NNLO predictions. The ratio $\RttDY$ is observed to be independent of the centrality, as expected.

\begin{table}[htbp]
\centering
\topcaption{
Measured \ttbar cross section, DY cross section, and $\RttDY$ ratio for three different \PbPb centrality selections. For each measurement, the statistical and systematic uncertainty components are shown separately.}
\label{tab:diff_summary}
\resizebox{\textwidth}{!}{
{\renewcommand{\arraystretch}{1.30}%
\begin{tabular}{lccc}
   Centrality bin & $\sigma_{\ttbar}\,[\mu\text{b}]$ & $\sigma_{\text{DY}}\,[\mu\text{b}]$ & $\RttDY$ \\[\cmsTabSkip]
   \hline
Inclusive & \sigmaTTbarobs & \sigmaDYobs & \RTTbarDYobs \\
0--10 \% & $1.49 ^{+0.40}_{-0.35} \stat^{+0.31}_{-0.25} \syst$ & $173.0 \pm 1.7 \stat^{+13.6}_{-12.7} \syst$ & $0.0082 ^{+0.0023}_{-0.0020} \stat^{+0.0017}_{-0.0014} \syst$ \\
10--90 \% & $1.90 ^{+0.40}_{-0.35} \stat^{+0.32}_{-0.27} \syst$ & $228.6 \pm 1.8 \stat^{+15.7}_{-14.8} \syst$ & $0.0083 ^{+0.0017}_{-0.0015} \stat^{+0.0012}_{-0.0011} \syst$ \\[\cmsTabSkip]
\end{tabular}
}}
\end{table}

\clearpage
\section{Summary}
\label{sec:conclusions}

The first measurement of the inclusive production of top quark pairs ($\ttbar$) in lead-lead (\PbPb) collisions at a nucleon-nucleon center-of-mass energy of $\sqrtsNN=5.36\TeV$ has been presented, extending earlier studies performed at lower collision energies. The analysis is based on dilepton events selected from the 2023 dataset recorded by the CMS experiment, corresponding to an integrated luminosity of 1.58\nbinv. Several improvements have been implemented---including enhanced electron identification and lepton isolation, refined heavy-flavor jet identification, and a dedicated calibration of the \PQb jet energy scale---which collectively provide a substantial gain in precision compared with previous analyses.

The cross sections $\sigma_{\ttbar}=\sigmaTTbarobs\mub$, and $\sigma_\text{DY}=\sigmaDYobs\mub$ for the Drell--Yan process, along with their ratio $\RttDY=\RTTbarDYobs$, have been measured. The $\sigma_{\ttbar}$ and $\RttDY$ measurements have total uncertainties of ${\approx}20\%$ and are consistent with perturbative quantum chromodynamics calculations at next-to-next-to-leading-order accuracy obtained using different nuclear parton distribution functions. Results for central and semicentral \PbPb collisions are also reported for the first time to examine the dependence of the top quark yields on the collision impact parameter. Combined with the higher collision energy and the larger data samples anticipated for the remainder of Run~3, these developments firmly establish top quark production as a future probe of initial- and final-state effects in heavy ions.

\begin{acknowledgments}
We congratulate our colleagues in the CERN accelerator departments for the excellent performance of the LHC and thank the technical and administrative staffs at CERN and at other CMS institutes for their contributions to the success of the CMS effort. In addition, we gratefully acknowledge the computing centers and personnel of the Worldwide LHC Computing Grid and other centers for delivering so effectively the computing infrastructure essential to our analyses. Finally, we acknowledge the enduring support for the construction and operation of the LHC, the CMS detector, and the supporting computing infrastructure provided by the following funding agencies: SC (Armenia), BMBWF and FWF (Austria); FNRS and FWO (Belgium); CNPq, CAPES, FAPERJ, FAPERGS, and FAPESP (Brazil); MES and BNSF (Bulgaria); CERN; CAS, MoST, and NSFC (China); MINCIENCIAS (Colombia); MSES and CSF (Croatia); RIF (Cyprus); SENESCYT (Ecuador); ERC PRG and PSG, TARISTU24-TK10 and MoER TK202 (Estonia); Academy of Finland, MEC, and HIP (Finland); CEA and CNRS/IN2P3 (France); SRNSF (Georgia); BMFTR, DFG, and HGF (Germany); GSRI (Greece); MATE and NKFIH (Hungary); DAE and DST (India); IPM (Iran); SFI (Ireland); INFN (Italy); MSIT and NRF (Republic of Korea); MES (Latvia); LMTLT (Lithuania); MOE and UM (Malaysia); BUAP, CINVESTAV, CONACYT, LNS, SEP, and UASLP-FAI (Mexico); MOS (Montenegro); MBIE (New Zealand); PAEC (Pakistan); MSHE, NSC, and NAWA (Poland); FCT (Portugal); MESTD (Serbia); MICIU/AEI and PCTI (Spain); MOSTR (Sri Lanka); Swiss Funding Agencies (Switzerland); MST (Taipei); MHESI (Thailand); TUBITAK and TENMAK (T\"{u}rkiye); NASU (Ukraine); STFC (United Kingdom); DOE and NSF (USA).

\hyphenation{Rachada-pisek} Individuals have received support from the Marie-Curie program and the European Research Council and Horizon 2020 Grant, contract Nos.\ 675440, 724704, 752730, 758316, 765710, 824093, 101115353, 101002207, 101001205, and COST Action CA16108 (European Union); the Leventis Foundation; the Alfred P.\ Sloan Foundation; the Alexander von Humboldt Foundation; the Science Committee, project no. 22rl-037 (Armenia); the Fonds pour la Formation \`a la Recherche dans l'Industrie et dans l'Agriculture (FRIA) and Fonds voor Wetenschappelijk Onderzoek contract No. 1228724N (Belgium); the Beijing Municipal Science \& Technology Commission, No. Z191100007219010, the Fundamental Research Funds for the Central Universities, the Ministry of Science and Technology of China under Grant No. 2023YFA1605804, the Natural Science Foundation of China under Grant No. 12535004, and USTC Research Funds of the Double First-Class Initiative No.\ YD2030002017 (China); the Ministry of Education, Youth and Sports (MEYS) of the Czech Republic; the Shota Rustaveli National Science Foundation, grant FR-22-985 (Georgia); the Deutsche Forschungsgemeinschaft (DFG), among others, under Germany's Excellence Strategy -- EXC 2121 ``Quantum Universe" -- 390833306, and under project number 400140256 - GRK2497; the Hellenic Foundation for Research and Innovation (HFRI), Project Number 2288 (Greece); the Hungarian Academy of Sciences, the New National Excellence Program - \'UNKP, the NKFIH research grants K 131991, K 138136, K 143460, K 143477, K 147557, K 146913, K 146914, K 147048, TKP2021-NKTA-64, and 2025-1.1.5-NEMZ\_KI-2025-00004, and MATE KKP and KKPCs Research Excellence and Flagship Research Groups grants (Hungary); the Council of Science and Industrial Research, India; ICSC -- National Research Center for High Performance Computing, Big Data and Quantum Computing, FAIR -- Future Artificial Intelligence Research, and CUP I53D23001070006 (Mission 4 Component 1), funded by the NextGenerationEU program, the Italian Ministry of University and Research (MUR) under Bando PRIN 2022 -- CUP I53C24002390006, PRIN PRIMULA 2022RBYK7T (Italy); the Latvian Council of Science; the Ministry of Science and Higher Education, project no. 2022/WK/14, and the National Science Center, contracts Opus 2021/41/B/ST2/01369, 2021/43/B/ST2/01552, 2023/49/B/ST2/03273, and the NAWA contract BPN/PPO/2021/1/00011 (Poland); the Funda\c{c}\~ao para a Ci\^encia e a Tecnologia (Portugal); the National Priorities Research Program by Qatar National Research Fund; MICIU/AEI/10.13039/501100011033, ERDF/EU, ``European Union NextGenerationEU/PRTR", projects PID2022-142604OB-C21, PID2022-139519OB-C21, PID2023-147706NB-I00, PID2023-148896NB-I00, PID2023-146983NB-I00, PID2023-147115NB-I00, PID2023-148418NB-C41, PID2023-148418NB-C42, PID2023-148418NB-C43, PID2023-148418NB-C44, PID2024-158190NB-C22, RYC2021-033305-I, RYC2024-048719-I, CNS2023-144781, CNS2024-154769 and Plan de Ciencia, Tecnolog{\'i}a e Innovaci{\'o}n de Asturias, Spain; the Chulalongkorn Academic into Its 2nd Century Project Advancement Project, the National Science, Research and Innovation Fund program IND\_FF\_68\_369\_2300\_097, and the Program Management Unit for Human Resources \& Institutional Development, Research and Innovation, grant B39G680009 (Thailand); the Eric \& Wendy Schmidt Fund for Strategic Innovation through the CERN Next Generation Triggers project under grant agreement number SIF-2023-004; the Kavli Foundation; the Nvidia Corporation; the SuperMicro Corporation; the Welch Foundation, contract C-1845; and the Weston Havens Foundation (USA).
\end{acknowledgments}

\bibliography{auto_generated}
\cleardoublepage \appendix\section{The CMS Collaboration \label{app:collab}}\begin{sloppypar}\hyphenpenalty=5000\widowpenalty=500\clubpenalty=5000\input{HIN-24-021-public-authorlist.tex}\end{sloppypar}
\end{document}

%% file: HIN-24-021-public-authorlist.tex
\cmsinstitute{Yerevan Physics Institute, Yerevan, Armenia}
{\tolerance=6000
A.~Belyaev\cmsorcid{0000-0003-1692-1173}, A.~Gevorgyan\cmsorcid{0000-0003-2751-9489}, A.~Hayrapetyan, A.~Tumasyan\cmsAuthorMark{1}\cmsorcid{0009-0000-0684-6742}
\par}
\cmsinstitute{Marietta Blau Institute for Particle Physics, Vienna, Austria}
{\tolerance=6000
W.~Adam\cmsorcid{0000-0001-9099-4341}, L.~Benato\cmsorcid{0000-0001-5135-7489}, T.~Bergauer\cmsorcid{0000-0002-5786-0293}, M.~Dragicevic\cmsorcid{0000-0003-1967-6783}, P.S.~Hussain\cmsorcid{0000-0002-4825-5278}, M.~Jeitler\cmsAuthorMark{2}\cmsorcid{0000-0002-5141-9560}, N.~Krammer\cmsorcid{0000-0002-0548-0985}, A.~Li\cmsorcid{0000-0002-4547-116X}, D.~Liko\cmsorcid{0000-0002-3380-473X}, M.~Matthewman, J.~Schieck\cmsAuthorMark{2}\cmsorcid{0000-0002-1058-8093}, R.~Sch\"{o}fbeck\cmsAuthorMark{2}\cmsorcid{0000-0002-2332-8784}, M.~Shooshtari\cmsorcid{0009-0004-8882-4887}, M.~Sonawane\cmsorcid{0000-0003-0510-7010}, N.~Van~Den~Bossche\cmsorcid{0000-0003-2973-4991}, W.~Waltenberger\cmsorcid{0000-0002-6215-7228}, C.-E.~Wulz\cmsAuthorMark{2}\cmsorcid{0000-0001-9226-5812}
% ...
\par}
\vskip\cmsinstskip
\dag:~Deceased\\
$^{1}$Also at Yerevan State University, Yerevan, Armenia\\
$^{2}$Also at TU Wien, Vienna, Austria\\
% ...
